\documentclass[letterpaper,11pt]{article}
\usepackage{jheppub}
\pdfoutput=1
\usepackage{amsmath,amssymb,amsfonts,bm,amscd,mathtools}
\usepackage{graphicx}
\usepackage{multirow}
\usepackage{verbatim}
\usepackage{appendix}
\usepackage{fancybox}
\usepackage{array}
\usepackage{url}
\usepackage{float}
\usepackage{subfigure}
\usepackage{tikz}
\usepackage{dsfont}

\def\CH{{\mathcal H}}

\def\be{\begin{equation}}
\def\ee{\end{equation}}
\def\bea{\begin{eqnarray}}
\def\eea{\end{eqnarray}}

\newcommand{\cH}{{\mathcal H}}
\newcommand{\tq}{{\mathtt q}}

\newcommand{\2}{{\mathtt 2}}
\newcommand{\3}{{\mathtt 3}}
\newcommand{\4}{{\mathtt 4}}
\newcommand{\ta}{{\mathtt a}}
\newcommand{\tb}{{\mathtt b}}

\newcommand{\scg}{{\psi \mathtt{CG}}}
\newcommand{\rcg}{{\rho \mathtt{CG}}}
\newcommand{\rpt}{{\rho \mathtt{PT}}}

\newcommand{\sZ}{{Z}}

\newcommand{\1}{{1}}

\title{Monotonicity conjecture for multi-party entanglement I}

\author{Abhijit Gadde$^{\,a}$, Shraiyance Jain$^{\,a}$, Vineeth Krishna$^{\,a}$, Harshal Kulkarni$^{\,a,b}$, Trakshu Sharma$^{\,a}$}
\affiliation[a]{Department of Theoretical Physics \\ 
Tata Institute for Fundamental Research, Mumbai 400005}
\affiliation[b]{Indian Institute of Science Education and Research, 
Kolkata 741246}
\emailAdd{abhijit@theory.tifr.res.in, shraiyance.jain@tifr.res.in, vineeth@theory.tifr.res.in, harshalkulkarni20@gmail.com, trakshu.sharma@tifr.res.in}

\abstract{In this paper, we conjecture a monotonicity property that we call monotonicity under coarse-graining for a class of multi-partite entanglement measures. We check these properties by computing the measures for various types of states using different methods.}

\begin{document}
\maketitle
\flushbottom

\section{Introduction and motivation}\label{intro}

What is a good measure of quantum entanglement? This question has been widely addressed for \emph{mixed} states, see \cite{Horodecki:2009zz} for a review and an extensive list of references. It is known that a good entanglement measure ${\cal M}(\rho_{A_\1,\ldots, A_\tq})$ defined for a density matrix over $\tq$ parties $A_\1$ to $A_\tq$  must have the following properties \cite{Bennett:1996gf, Vedral:1997qn, Vidal:1998re},
\begin{enumerate}
    \item ${\cal M}(\rho_{A_\1,\ldots, A_\tq})$ is invariant under local unitary transformations ({\tt lu}) i.e. under,
    \begin{align}\label{lu}
        \rho_{A_\1,\ldots, A_\tq}\to U^\dagger_{\1}\otimes\ldots \otimes  U^\dagger_{\tq}\, \rho_{A_\1, \ldots, A_\tq}\, U_{\1}\otimes \ldots \otimes  U_{\tq}.
    \end{align}
    \item ${\cal M}(\rho_{A_\1,\ldots, A_\tq})$ is monotonically non-increasing  under local operations and classical communications ({\tt locc}) i.e. under,
    \begin{align}\label{locc}
        \rho_{A_\1,\ldots, A_\tq}\to \sum_{i} L^{(i)\dagger}_{\1}\otimes\ldots \otimes  L^{(i)\dagger}_{\tq}\, \rho_{A_\1, \ldots, A_\tq}\, L^{(i)}_{\1}\otimes \ldots \otimes  L^{(i)}_{\tq}.
    \end{align}
    where the operators $L^{(i)}_\ta$ obey the trace preserving condition $\sum_i L^{\dagger (i)}_\ta L^{(i)}_\ta={\mathbb I}$ and the overall operation is also trace preserving i.e. $\sum_i \otimes_\ta L^{\dagger (i)}_\ta L^{(i)}_\ta={\mathbb I}$.
\end{enumerate}
The rationale for the first condition is clear. The local unitary operations furnish a change of basis in individual Hilbert spaces and it is expected that the entanglement measure is invariant under such a change of basis. The second condition requires a little more explanation. 

The trace preserving operation $\rho \to \sum_i L_\ta^{(i)\dagger}\rho L_\ta^{(i)}$ for a fixed party $A_{\ta}$ can be implemented by a unitary acting on $H_{A_\ta}\otimes H_{\rm env}$ and then tracing over $H_{\rm env}$. This operation is known as a local operation ({\tt lo}) or as a quantum channel. In this context, the operators $A_{\ta}^{(i)}$ are known as the Kraus operators. An example of such an operation is a measurement of some observable. 
A more general operation of the kind given in \eqref{locc} comprises of local operations on multiple parties, but not just that,  the overall sum over $i$ further correlates these operations. This correlation models classical communication i.e. a (classical) phone call to the other party to perform correlated local operations. Together the operation \eqref{locc} is known as local operation and classical communication ({\tt locc}). As the system interacts with the environment during {\tt locc}, some of the entanglement of the original state is now shared with the environment. That is why it is expected that an $\tt locc$ operation must decrease quantum correlations. It may lead to new classical correlations however owing to the phone call. The first condition, {\tt lu} invariance, can also be understood in view of the second condition. A local unitary transformation is an invertible {\tt locc} operation and given that the measure must be non-increasing under {\tt locc}, it must be constant under any local unitary transformation.

A corollary of monotonicity under {\tt locc} is that the measure ${\cal M}(\rho_{A_\1,\ldots, A_\tq})=0$ for completely separable mixed states i.e. for the density matrices that take the form
\begin{align}
    \rho_{A_\1,\ldots, A_\tq} = \sum_i \,p_i\, \rho_{A_\1}^{(i)}\otimes \ldots \otimes \rho_{A_{\tq}}^{(i)}.
\end{align}
where $p_i$ are probabilities and $\rho_{A_{\ta}}^{(i)}$ are some density matrices for party $A_\ta$. This is understood as follows. The completely separable states have the property that {\tt locc} operations are reversible on them \cite{Vidal:1998re}. So the entanglement measure must be a constant on them. We can subtract this constant from the definition of ${\cal M}$ and make the measure have value $0$ on the completely separable states. This is interpreted as completely separable states having only classical correlation and no quantum correlations, see section XV.B.2 in \cite{Horodecki:2009zz}.

\subsection*{Pure states}
We are interested in understanding the multi-partite entanglement structure of states of quantum field theory and gravity.  Naturally, it's the \emph{pure} states that take center stage. We will only restrict our discussion of the multi-partite entanglement measure to pure states and will not attempt to extend it to mixed states.\footnote{There is a standard way to extend the measure defined on the pure states to mixed states via the so-called ``convex roof construction'' \cite{Uhlmann1998,Vidal:1998re}. See section \ref{discuss}. \label{foot1}}  Because the property of monotonicity under {\tt locc} is defined inherently only for mixed states, it will not play any role in the rest of the paper.  In addition to demanding that the measure ${\cal M}$ for pure states be invariant under local unitary transformations, we will also require certain other appealing properties. 

\begin{figure}[t]
    \begin{center}
        \includegraphics[scale=0.4]{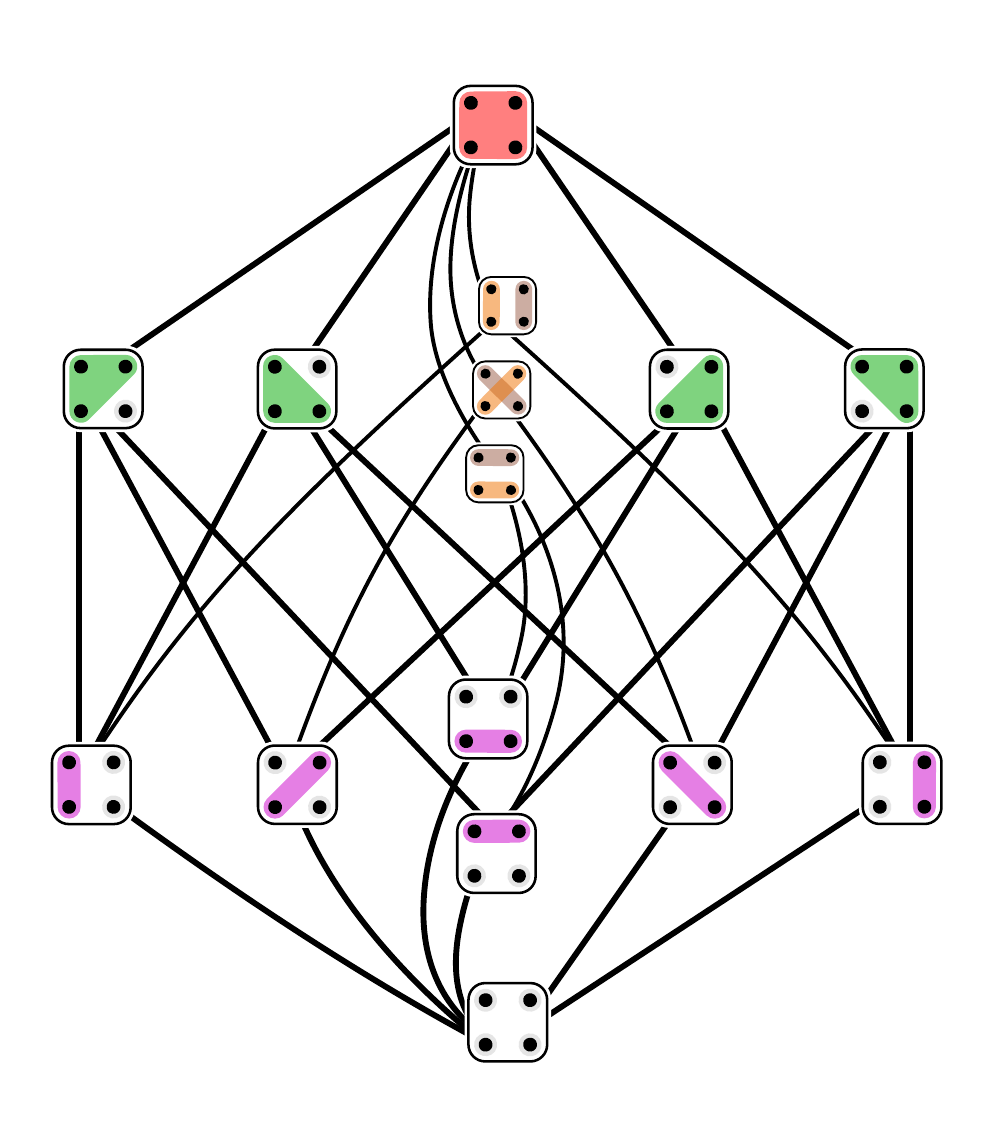}
    \end{center}
    \caption{Hasse diagram for partial order under set refinement for the set of four elements \cite{wiki}.}\label{Set_partitions_4}
\end{figure}

\begin{itemize}
    \item ${\cal M}^{(\tq)}(|\Psi\rangle_{A_\1,\ldots, A_\tq})$  is \emph{symmetric} in all the parties i.e. it is invariant under permutations of the parties.
    \begin{align}
        {\cal M}^{(\tq)}(|\Psi\rangle_{A_\1,\ldots, A_\tq}) = {\cal M}^{(\tq)}(|\Psi\rangle_{A_{\sigma \cdot\1},\ldots, A_{\sigma\cdot \tq}}),\qquad {\rm for}\quad \sigma\in S_{\tq}.
    \end{align}
    Here we have put the superscript $(\tq)$ to emphasize that the measure is defined on $\tq$-partite states.
    \item ${\cal M}^{(\tq)}(|\Psi \rangle \otimes |\Phi \rangle)$ for direct product states is the sum of their individual measures. More concretely, for the two $\tq$-partite states $|\Psi\rangle \in {\cal H}^{\Psi}_{A_\1}\otimes \ldots \otimes {\cal H}^{\Psi}_{A_\tq}$ and $|\Phi\rangle \in {\cal H}^{\Phi}_{A_\1}\otimes \ldots \otimes {\cal H}^{\Phi}_{A_\tq}$ consider the direct product state $|\Psi\rangle \otimes |\Phi\rangle$ also as a $\tq$-partite state valued in ${\cal H}_{A_\1}\otimes \ldots \otimes {\cal H}_{A_\tq}$ where ${\cal H}_{A_\ta}\equiv {\cal H}_{A_\ta}^{\Psi}\otimes {\cal H}_{A_\ta}^{\Phi}$. Then we require,
    \begin{align}
        {\cal M}^{(\tq)}(|\Psi\rangle \otimes |\Phi\rangle) = {\cal M}^{(\tq)}(|\Psi\rangle) + {\cal M}^{(\tq)}(|\Phi\rangle).
    \end{align}
\end{itemize}
Given that monotonicity under {\tt locc} is not at our disposal, we need an alternative notion of loss of entanglement and monotonicity under it. Let us assume that we have a family of multi-partite measures ${\cal M}^{(\tq)}$. In that case, we would like to propose the following condition that quantifies the loss of entanglement.  If  we identify  two parties and treat them as a single party and compute ${\cal M}^{(\tq-1)}$ on the resulting $\tq-1$ partite state then we require,
\begin{itemize}
    \item {\bf Monotonicity under coarse graining}:
    \begin{align}\label{psi-mono}
        {\cal M}^{(\tq)}(|\Psi\rangle)-{\cal M}^{(\tq-1)}(|\Psi_{[A_\ta A_\tb]}\rangle) \geq 0 \qquad \quad \forall \, \ta,\tb.
    \end{align}
\end{itemize}
We use the notation $|\Psi_{[A_\ta A_\tb]}\rangle$ to denote the $\tq-1$ partite state obtained after identification of parties $A_\ta$ and $A_{\tb}$.
Because the measure ${\cal M}$ is invariant under local unitary transformations, ${\cal M}^{(\tq-1)}(|\Psi_{[A_\ta A_\tb]}\rangle)$ is effectively a measure obtained after the action of an arbitrary scrambling unitary transformation on ${\cal H}_\ta\otimes {\cal H}_\tb$. It is reasonable to expect that such a scrambling unitary reduces the multi-partite entanglement. The condition of monotonicity under coarse graining is not novel. It has already been considered for multi-partite states in \cite{Hein_2004}. 
Because the monotonicity property \eqref{psi-mono} is expected for all values of $\tq$, it implies a broad set of inequalities on a $\tq$-partite state coming from sequentially identifying parties. 
More formally, given a set of parties $\{A_\1, \ldots, A_\tq\}$, consider a set of all of its partitions. There is a natural partial order on this set coming from splitting or refinement. An example of this partial order for the case of four parties is shown in figure \ref{Set_partitions_4} with the help of the so-called ``Hasse diagram''.
Coarse graining is the opposite of refinement where multiple partitions are combined to form a single partition.  The monotonicity under coarse graining asserts that the measure ${\cal M}$ respects this partial ordering. It means that the multi-partite measure is non-increasing as we follow a downward path in the Hasse diagram.
As ${\cal M}^{(\tq)}(|\Psi\rangle )-{\cal M}^{(\tq-1)}(|\Psi_{[A_\ta A_\tb]}\rangle )$ is non-negative, this difference can also be thought of as a multi-partite analogue of mutual-information between party $A_\ta$ and $A_\tb$. 

\subsection{For density matrices}
The monotonicity condition \eqref{psi-mono} for $\tq$-partite pure state $|\Psi\rangle$ can be easily reformulated for the density matrix $\rho$ on $\tq-1$ parties obtained after tracing out one of the parties, say party $\tq$. Note that here we are merely talking about the \emph{reformulation} of the pure state measure for density matrices and not its \emph{extension} to mixed states mentioned in footnote \ref{foot1}. We will denote the measures computed for the density matrix defined on $\tq-1$ parties with a superscript $(\tq)$ because the superscript stands for the number of parties in the description of the state as a pure state and any mixed state can be purified by introducing an extra party. Even though there are multiple ways to purify a density matrix, the measure ${\cal M}^{(\tq)}$ does not depend on the purification. This is because all purifications are related to each other by the action of local unitary transformations on the new party $A_\tq$ introduced by the purification and the measure ${\cal M}^{(\tq)}$ is invariant under it.
For the density matrix,  the monotonicity condition takes the form of two different conditions
\begin{itemize}
    \item {\bf Monotonicity under coarse graining}:
    
    Denoting the $\tq-2$ partite density matrix obtained by identifying parties $A_\ta$ and $A_\ta$ from $\rho$ as $\rho_{[A_\ta A_\tb]}$, 
    \begin{align}\label{rho-cg}
        {\cal M}_n^{(\tq)}(\rho) &\geq {\cal M}_n^{(\tq-1)}(\rho_{[A_\ta A_\tb]}).
    \end{align}
    \item {\bf Monotonicity under partial tracing}:
    
    Denoting the $\tq-2$ partite density matrix obtained by tracing out party $A_\ta$ from $\rho$ as $\rho_{[A_\ta]}$, 
    \begin{align}\label{rho-pt}
        {\cal M}_n^{(\tq)}(\rho) &\geq {\cal M}_n^{(\tq-1)}(\rho_{[A_\ta]}).
    \end{align}
    For the original $\tq$-partite pure state, this corresponds to the identification of $A_\ta$ with $A_\tq$ that is traced out to produce the density matrix.
\end{itemize}  

\subsection*{The conjecture}
The following conjecture is the main subject of this paper.
\begin{itemize}
    \item Renyi multi-entropy  \cite{Gadde:2022cqi, Gadde:2023zzj} $S_n^{(\tq)}$, for all positive integers $n$, is monotonic under coarse graining. 
\end{itemize}
In the rest of the paper, we will give support for this conjecture by considering various types of multi-partite pure states. 

\subsection*{Outline of the paper}
In the rest of the section \ref{intro}, we define the Renyi multi-entropy after introducing a method to construct general local unitary invariants using permutations and formulate the monotonicity conjecture precisely. We then verify it for certain simple states as a sanity check. In section \ref{sec-factorized}, we compute the Renyi multi-entropy for states that are in the neighborhood of the factorized state and verify the conjecture for them. We then prove the conjecture for classical probability distributions in section \ref{probs}. From the quantum perspective, classical probability distributions are thought of as diagonal density matrices. The conjecture is then proved for a generalized version of the so-called W state on three parties in section \ref{genw}. This example is interesting because the computation of $S_n^{(\tq)}$ reduces to the computation of the partition function of a dimer model on a periodic square lattice. This model is integrable and its partition function is computed using the techniques developed in \cite{KASTELEYN19611209}. In section \ref{holo}, we consider the monotonicity conjecture for holographic states and verify it using the prescription of \cite{Gadde:2022cqi,Gadde:2023zzj}. We provide the details of numerical checks of the conjecture for finite-dimensional multi-partite states chosen randomly in section \ref{numerical} and end with the summary and outlook. The paper is supplemented by one appendix describing Kasteleyn's approach to dimer models.

\subsection{Renyi multi-entropy: definition}\label{mrenyi-def}
Consider a normalized quantum state $|\Psi\rangle \in \otimes_{\ta=1}^\tq {\cal H}_\ta$ of a $\tq$-party system. Let $d_\ta$ be the dimension of the Hilbert space $\cH_\ta$ and $|\alpha_\ta\rangle, \, \alpha_\ta=1,\ldots, d_\ta$ be a set of its orthonormal basis vectors. In $|\alpha_\ta\rangle$ basis, the state $|\Psi\rangle$ is given as
\begin{align}
    |\Psi\rangle=\sum_{\alpha_1=1}^{d_1}\ldots \sum_{\alpha_\tq=1}^{d_\tq}  \,\,\psi_{\alpha_1\ldots \alpha_\tq}\, \,|\alpha_1\rangle\otimes \ldots \otimes|\alpha_\tq\rangle.
\end{align}
We call $\psi_{\alpha_1\ldots\alpha_\tq}$, the wavefunction.
Local unitary transformations take the form $|\Psi\rangle \to \otimes_{\ta=1}^\tq {\cal U}_\ta|\Psi\rangle$ where ${\cal U}_\ta$ is the unitary transformation that acts on ${\CH_\ta}$.  
Naturally, the local unitary invariants are constructed by taking multiple copies of $\psi$'s and $\bar \psi$'s and contracting the fundamental indices $\alpha_\ta$ of $\psi$ with anti-fundamental indices $\alpha_\ta$ of $\bar \psi$. As a result, the number of $\psi$'s and the number of $\bar \psi$'s is the same in any entanglement measure. We call this number the replica number $m$. We index the replicas by the superscript $(i)$. The Hilbert space $\CH_\ta$ of the $i$-th replica is denoted as $\CH_\ta^{(i)}$ and its basis as $|\alpha_\ta^{(i)}\rangle$. The wavefunction of the $i$-th replica is then $\psi_{\alpha_1^{(i)}\ldots \alpha_\tq^{(i)}}$ and its conjugate is $\bar \psi^{\alpha_1^{(i)}\ldots \alpha_\tq^{(i)}}$. A general invariant $\cal E$ can be written in terms of contractions of fundamental indices of $m$-replicas of $\psi$ with anti-fundamental indices of $m$-replicas of $\bar \psi$  as follows
\begin{align}
    {\cal E}                                                & =\Big(\psi_{\alpha_1^{(1)}\ldots \alpha_\tq^{(1)}}\ldots \psi_{\alpha_1^{(m)}\ldots \alpha_\tq^{(m)}}\Big)\Big( \bar \psi^{\beta_1^{(1)}\ldots \beta_\tq^{(1)}}\ldots \bar \psi^{\beta_1^{(m)}\ldots \beta_\tq^{(m)}} \Big)\delta^{\vec \alpha_1}_{\sigma_1\cdot\vec\beta_1}\ldots \delta^{\vec \alpha_\tq}_{\sigma_\tq\cdot\vec\beta_\tq}\notag \\
    {\rm where}\quad \delta^{\vec \alpha_\ta}_{\sigma_\ta\cdot\vec\beta_\ta} & \equiv
    \delta^{\alpha_\ta^{(1)}}_{\beta_\ta^{(\sigma_\ta\cdot 1)}}\ldots \delta^{\alpha_\ta^{(m)}}_{\beta_\ta^{(\sigma_\ta\cdot m)}}
\end{align}
If we call the pair of $\psi$ and $\bar \psi$ as a replica then the invariant  ${\cal E}$ is labeled by $\tq$ permutation elements $\sigma_\ta$ of the permutation group $S_m$ acting on the replica set. The element $\sigma_\ta$ indicates how the $m$ fundamental indices of party $\ta$ are contracted with $m$ anti-fundamental indices. See \cite{Gadde:2023zzj} for a more detailed discussion of general $\tq$-partite invariants. 

The Renyi multi-entropy is now defined starting from an invariant ${\cal E}_n^{(\tq)}$ labeled by a specific set of permutation elements $\sigma_\1,\ldots, \sigma_\tq$. Consider $m=n^{\tq-1}$ replicas, arranged as a $\tq-1$ dimensional regular hyper-cubical lattice. Let us label these directions by $\1,\ldots ,\tq-1$ and the ${\mathbb Z}_n$ discrete ``translations'' acting along each direction as ${\mathbb Z}_{n,\ta}$. 
The permutation $\sigma_\tq$ is taken to be identity and the rest of the permutations are taken to be the generators of independent ${\mathbb Z}_n$ symmetries i.e. $\sigma_{\ta}=g_{\ta}$ for $\ta=\1,\ldots, \tq-1$, where $g_{\ta}$ generates ${\mathbb Z}_{n,\ta}$. Although it might seem that $\tq$-th party is being treated differently from the rest of the parties, the invariant ${\cal E}_n^{(\tq)}$ is symmetric in all the parties. 
The Renyi multi-entropy is now defined as 
\begin{align}\label{m-renyi-def}
    S_{n}^{(\tq)}:=\frac{1}{1-n}\frac{1}{n^{\tq-2}} {\rm log}({\cal E}_n^{(\tq)}).
\end{align}
It is clear that $S_{n}^{(2)}$ is the ordinary Renyi-entropy $S_n$. It follows from the definition, that the measures ${\cal E}_n^{(\tq)}$ and  $S_n^{(\tq)}$ enjoy homogeneity in the state respectively i.e.,
\begin{align}\label{linear}
    {\cal E}_n^{(\tq)}(|\Psi_1\rangle \otimes |\Psi_2\rangle)&={\cal E}_n^{(\tq)}(|\Psi_1\rangle)\cdot {\cal E}_n^{(\tq)}(|\Psi_2\rangle)\notag\\
    S_n^{(\tq)}(|\Psi_1\rangle \otimes |\Psi_2\rangle)&=S_n^{(\tq)}(|\Psi_1\rangle)+S_n^{(\tq)}(|\Psi_2\rangle)
\end{align}
for any $\tq$-partite states $|\Psi_1\rangle, |\Psi_2\rangle$.

In many cases, the monotonicity conjecture for the Renyi multi-entropy is more conveniently thought of as the inequality for ${\cal E}_n^{(\tq)}$ before taking the logarithm,\footnote{Note that in defining the Renyi multi-entropy $S_n^{(\tq)}$ in equation \eqref{m-renyi-def}, we have tacitly assumed that the quantity ${\cal E}_n^{(\tq)}$ is positive. For this paper, this is not an essential assumption. The monotonicity conjecture can equally well be expressed in terms of ${\cal E}_n^{(\tq)}$ as done in equation \eqref{E-ineq}, \eqref{E-rho-cg} and \eqref{E-rho-pt}.}
\begin{align}\label{E-ineq}
    {S}_n^{(\tq)}(|\psi\rangle )\geq {S}_n^{(\tq-1)}(|\psi_{[A_\ta A_\tb]}\rangle)\quad
    \Leftrightarrow \quad {\cal E}_n^{(\tq)}(|\psi\rangle )\leq \Big({\cal E}_n^{(\tq-1)}(|\psi_{[A_\ta A_\tb]}\rangle)\Big)^n.
\end{align}
Equivalently, following \eqref{rho-cg} and \eqref{rho-pt},  we have two inequality conjectures for ${\cal E}$ in terms of density matrices 
\begin{align}
    {S}_n^{(\tq)}(\rho )&\geq {S}_n^{(\tq-1)}(\rho_{[A_\ta A_\tb]})\hspace{-2.3cm}  &\Leftrightarrow\quad  {\cal E}_n^{(\tq)}(\rho )&\leq \Big({\cal E}_n^{(\tq-1)}(\rho_{[A_\ta A_\tb]})\Big)^n, \label{E-rho-cg}\\
    {S}_n^{(\tq)}(\rho)&\geq {S}_n^{(\tq-1)}(\rho_{[A_\ta]})\hspace{-2.3cm} &\Leftrightarrow \quad{\cal E}_n^{(\tq)}(\rho )&\leq \Big({\cal E}_n^{(\tq-1)}(\rho_{[A_\ta]})\Big)^n.\label{E-rho-pt}
\end{align}
Equation \eqref{E-rho-cg} corresponds to monotonicity under coarse graining and equation \eqref{E-rho-pt} corresponds to monotonicity under partial trace. As we will refer to these inequalities frequently in the bulk of the paper, we will term the inequalities \eqref{E-ineq}, \eqref{E-rho-cg} and \eqref{E-rho-pt} as $\psi {\mathtt{CG}}$, $\rho {\mathtt {CG}}$ and $\rho {\mathtt {PT}}$ respectively ($\mathtt{CG}$ and ${\mathtt{PT}}$ stand for coarse graining and partial trace respectively).
In order to understand these conjectures better, it is useful to ask when inequality $\scg$ is saturated. 
Even in the simplest non-trivial case of $\tq=3$, we can not characterize the saturation in complete generality but in certain cases, as we will see in the main text of the paper, its saturation corresponds to the vanishing of mutual information between the parties being identified. We numerically find that the saturation of $\scg$ and vanishing of mutual information are indeed correlated for general tri-partite states in a small number of dimensions. We find it reasonable to conjecture that for tri-partite states the inequality  $\scg$ is saturated if and only if mutual information between party $A_\ta$ and $A_\tb$ vanishes. For higher partite states, however, we do not find any correlation between the saturation of $\scg$ and the vanishing of mutual information. 

At this point, it is also worth noting that the conjectured inequalities are between the two quantities ${\cal E}_n^{(\tq)}$ and $({\cal E}_n^{\tq-1})^n$ that have the same homogeneity in $\psi$. So in establishing these inequalities, the normalization of the state does not matter. The state $|\Psi\rangle$ can have any norm as long as it is the same on both sides of the inequality.  

The Renyi multi-entropy $S_n^{(\tq)}$ was introduced in \cite{Gadde:2022cqi} mainly due to its applications to holographic states. For $\tq = 2$, it reduces to the usual Renyi entropy $S_n$ and its $n \to 1$ limit reduces to the usual Von Neumann entropy, both of which have holographic interpretations \cite{Headrick:2010zt, Hung:2011nu, Dong:2016fnf, Ryu:2006bv, Casini:2011kv, Lewkowycz:2013nqa, Faulkner:2013ana, Engelhardt:2014gca}. Consideration of $S_n^{(\tq)}$ is then generalizing these concepts to the case of multi-partite systems. It turns out, at least for some range of values for $n$, that it is computed by a geometric quantity on the bulk side. See section \ref{holo} for the discussion of $S_n^{(\tq)}$ for the holographic states.

\subsubsection{Density matrix and lattices}\label{lattice}
\begin{figure}[t]
    \begin{center}
        \includegraphics[scale=1.0]{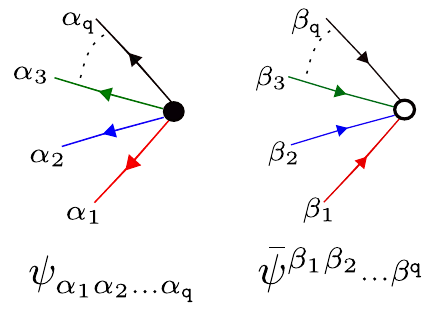}
    \end{center}
    \caption{Graphical notation for the wavefunction $\psi$ and its conjugate $\bar \psi$.}\label{psi-barpsi}
\end{figure}

To discuss the measure ${\cal E}_n^{(\tq)}$ more concretely, it is convenient to introduce a graphical notation in which the wavefunction $\psi$ is denoted as a $\tq$-valent black vertex as in figure \ref{psi-barpsi}.
The vertex has outgoing colored edges as they stand for fundamental indices $\alpha_\ta$ of distinct Hilbert spaces $\CH_\ta$. Similarly, $\bar \psi$ is denoted as a $\tq$-valent white vertex with incoming colored edges. The index contraction in $\CH_\ta$ is denoted as joining a white vertex with a black vertex with the appropriately colored edge. An entanglement measure $\cal E$ is then a bi-partite graph made out of these vertices with no dangling edges. 
\begin{figure}[t]\label{S3-rho}
    \begin{center}
        \includegraphics[scale=0.8]{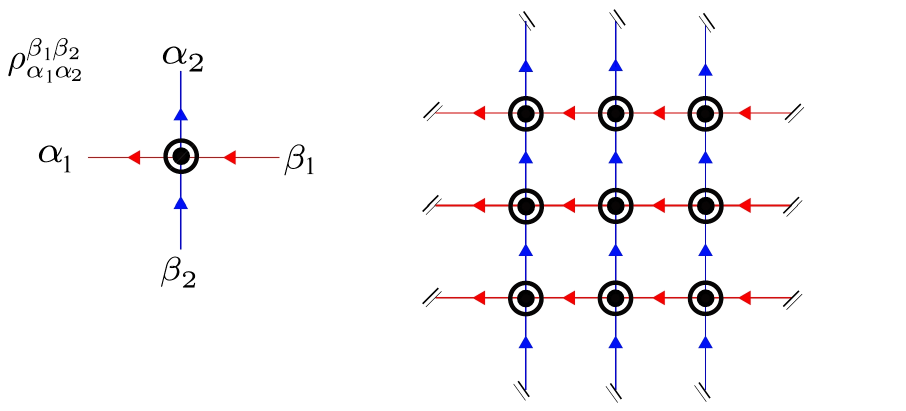}
    \end{center}
    \caption{A single connected component of the measure corresponding to choosing $\sigma_1$, $\sigma_2$ and $\sigma_3$ to be three independent generators of ${\mathbb Z}_3\times {\mathbb Z}_3\times {\mathbb Z}_3$.}\label{q3}
\end{figure}

It is more convenient to think about ${\cal E}_n^{(\tq)}$ in terms of density matrices because such a description admits a statistical mechanical interpretation. The density matrix obtained by tracing out the $\tq$'th party is denoted as a black-white vertex with $\tq-1$ number of pairs of ingoing and outgoing colored arrows. These arrows correspond to the fundamental and anti-fundamental indices of the remaining $\tq-1$ parties.  It is particularly convenient to arrange a pair of arrows of a given color along a  basis vector of an orthogonal frame in $\tq-1$ dimensions that is oriented in the opposite direction. A graphical notation for the density matrix on two parties is given in figure \ref{q3}. In graphical presentation, a density matrix resembles a lattice point in a $\tq-1$ dimensional hyper-cubic lattice. The contractions of indices in  ${\cal E}_n^{(\tq-1)}(\rho)$  corresponds to an arrangement of $\rho$'s in $\tq-1$ dimensional hyper-cubic lattice of size $n\times \ldots \times n$ with periodic boundary conditions. This is illustrated in figure \ref{q3} for ${\cal E}_{3}^{(\3)}$. In this description, the quantity ${\cal E}_n^{(\tq)}$ is thought of as the partition function of a statistical mechanical model on this lattice with the ``Boltzmann weight'' of each vertex given by $\rho_{A_1,\ldots, A_{\tq-1}}$. Because the measure is symmetric in all parties, tracing out another party instead of $A_{\tq-1}$ party gives another dual description of the same lattice model. There are $\tq$ equivalent descriptions of a given lattice model depending on which party we trace out to construct the density matrix on the rest of the parties. 

The quantity $({\cal E}_{n}^{(\tq-1)}(\rho_{[A_\ta]}))^n$ 
appearing on the right-hand side of inequality \eqref{rho-pt} has a nice interpretation from the lattice point of view. It is the partition function on the lattice with the same set of vertices but with bonds in the $\ta$ direction deleted. As a result, the lattice disconnects in the $\ta$ direction and instead becomes $n$ disconnected $\tq-2$ dimensional lattice ``layers'' in the transverse directions. The Boltzmann weight of the new vertex is given by the density matrix $\rho_{[A_\ta]}$ obtained by tracing party $A_\ta$. The quantity $({\cal E}_{n}^{(\tq-1)}(\rho_{[A_\ta A_\tb]}))^n$ appearing on the right hand side of inequality \eqref{rho-cg} also has an  interpretation in terms of the lattice. It is the partition function on the lattice with the same set of vertices but with bonds in the $\ta$ direction disconnecting and going instead in the $\tb$ direction forming ``double bonds'' with the already existing bonds in the $\tb$ direction. As a result, again the lattice disconnects in the $\ta$ direction and instead becomes $n$ disconnected ``layers'' of $\tq-2$ dimensional lattice in the transverse directions. 

For future use, we observe the following. Let us pluck out one $\rho$ vertex from the lattice corresponding to ${\cal E}_n^{(\tq)}$. The remaining lattice yields a matrix $H_{\alpha_\1,\ldots,\alpha_{\tq-1}}^{\beta_\1,\ldots,\beta_{\tq-1}}(\rho) \equiv H_n^{(\tq)}(\rho)$ on $\tq-1$ parties such that ${\cal E}_n^{(\tq)}={\rm Tr}\, \rho H_n^{(\tq)}(\rho)$. Here the trace is taken over all the $\tq-1$ parties. We observe that the matrix $H_n^{(\tq)}(\rho)$ is hermitian. This is because the hermitian conjugate operation on any graph of $\rho$'s reverses all the arrows. If the graph reversed arrows is isomorphic (identical up to relabelling of vertices) to the original one then the corresponding quantity constructed out of $\rho$'s is hermitian. This is the case for $H_n^{(\tq)}(\rho)$.

\subsection{Certain simple states}\label{simple}
Let us get some feel for the monotonicity conjecture for certain simple quantum states. It is convenient to keep the hyper-cubic lattice model of  ${\cal E}_n^{(\tq)}$ in mind for these computations. 
\subsubsection{Factorized state}\label{factorized}
Perhaps the simplest quantum state is the one corresponding to a factorized density matrix,
\begin{align}
    \rho_{A_\1,\ldots, A_{\tq-1}}=\rho_{A_1}\otimes \ldots \otimes \rho_{A_{\tq-1}}.
\end{align}
In this case, the $\tq-1$ dimensional lattice breaks up into $n^{\tq-2}$ copies of linear $1$ dimensional lattices in each of the $\tq-1$ directions. We get
\begin{align}\label{m-renyi-factorized}
    {\cal E}_n^{(\tq)}&=\prod_\ta {\rm Tr}\rho_{A_\ta}^n.\notag\\
    S_n^{(\tq)}&=\frac{1}{n^{\tq-2}}\sum_\ta \, S_n^{(2)}(\rho_{A_\ta}).
\end{align}
It is clear that the Renyi multi-entropy of the $\tq-2$ partite state after tracing out $A_\tb$,  $S_n^{(\tq)}(\rho_{[A_\tb]})$, is given by the same expression as above but without the $\ta=\tb$ term. As the Renyi entropy $S_n^{(2)}$ is positive, the inequality $\rpt$ is obeyed. The measure after identification of some two parties $A_\ta$ and $A_\tb$, $S_n^{(\tq)}(\rho_{[A_\ta A_\tb]})$, on the other hand is the same as  $S_n^{(\tq)}(\rho)$. So the inequality $\rcg$ is saturated.

In fact, the inequality $\rcg$ is saturated for a more general type of density matrix, 
\begin{align}\label{rho-factor}
    \rho_{A_{\1},\ldots, A_{\tq-1}}= (\rho_{A'_{\ta}} \otimes {\mathbb I}_{A_\ta}) \otimes (\rho_{A'_{\tb}} \otimes {\mathbb I}_{A_\tb})
\end{align}
Here the density matrices $\rho_{A'_\ta}, \rho_{A'_\tb }$ are $\tq-2$ party density matrices that do not act on $A_\ta$ and $A_\tb$ respectively. In this case, thanks to equation \eqref{linear}, the measure ${\cal E}_n^{(\tq)}$ factorizes.
\begin{align}
    {\cal E}_n^{(\tq)}(\rho)&=  {\cal E}_n^{(\tq)}(\rho_{A'_\ta}\otimes {\mathbb I}_{A_\ta}) \,{\cal E}_n^{(\tq)}(\rho_{A'_\tb}\otimes {\mathbb I}_{A_\tb})= {\cal E}_n^{(\tq-1)}(\rho_{A'_\ta}) \,{\cal E}_n^{(\tq-1)}(\rho_{A'_\tb}).
\end{align}
From this expression, it is clear that both terms separately saturate the inequality $\rcg$. Interestingly, the density matrix \eqref{rho-factor}, for $\tq=3$ becomes the factorized density matrix $\rho_{A_\1,A_\2}=\rho_{A_\1}\otimes\rho_{A_\2}$. This is the density matrix for which the mutual information between party $A_\1$ and $A_\2$ vanishes. As we have seen, such a density matrix also saturates the inequality $\rcg$. However, for a higher number of parties, there may be density matrices apart from the type \eqref{rho-factor} that saturate the inequality $\rcg$ as well. So we can not say anything definitive about the relationship between the vanishing of mutual information and the vanishing of ``$\tq$-partite mutual information'' i.e. saturation of the inequality $\rcg$.

\subsubsection{Bi-partite states}
Let us consider a $\tq$-partite state which only has pairwise bi-partite entanglement. In other words, it is given by the tensor product
\begin{align}
    |\Psi\rangle = \bigotimes_{\ta=\1}^\tq  \bigotimes_{\tb=1; \tb>\ta}^\tq  |\Psi_{\ta \tb}\rangle.
\end{align} 
In writing this state we have assumed that any given party $A_\ta$ consists of the tensor product of $\tq-1$ Hilbert spaces each of which is entangled separately with each of the remaining $\ta-1$ parties.

Using equation \eqref{linear}, we get
\begin{align}\label{sumab}
    S_n^{(\tq)}(|\Psi\rangle) =\frac{1}{n^{\tq-2}} \sum_{\ta,\tb=1;\tb>\ta} S_n^{(2)}(|\Psi_{\ta\tb}\rangle) = \frac{1}{n^{\tq-2}}\frac{1}{2} \sum_{\ta=1}^\tq \Big(\sum_{\tb=1}^{\tq} S_n^{(2)}(|\Psi_{\ta\tb}\rangle)\Big).
\end{align}
The term in the bracket on the right-hand side is simply the Renyi entropy of party $A_\ta$. So we have
\begin{align}
    S_n^{(\tq)}(|\Psi\rangle)= \frac{1}{2} \frac{1}{n^{\tq-2}} \sum_{\ta=1}^\tq S_n^{(2)}(\rho_{A_\ta}).
\end{align}
When we compute the Renyi multi-entropy of the state obtained by identifying, say parties $A_\1$ and $A_\2$, we get the same expression as in equation \eqref{sumab} except that the term corresponding to the pair $(\ta,\tb)=(\1,\2)$ is absent. Because the Renyi entropies are positive, we have $\scg$ obeyed.

\subsubsection{Generalized GHZ state}
The Greenberger-Horne-Zeilinger (GHZ) state is the following entangled state of three qubits 
\begin{align}
    |{\mathrm{GHZ}}\rangle \equiv \frac{1}{\sqrt{2}}(|000\rangle+|111\rangle).
\end{align}
This state has played an important role in establishing the indispensable nature of quantum entanglement and is also used in quantum communication and cryptography. Inspired by the form of this state, we write a $\tq$-partite generalization of the  GHZ state. It is defined on $\tq$ parties each with  Hilbert space dimension $d$ as
\begin{align}
    |\Psi\rangle=\sum_{i=1}^d \lambda_i |e_{\1,i}\rangle \otimes\ldots \otimes |e_{\tq,i}\rangle,
\end{align}
where $|e_{\ta,i}\rangle$ forms an orthonormal basis of  party $A_\ta$. Normalizing the state sets $\sum_i|\lambda_i|^2=1$. To compute ${\cal E}_n^{(\tq)}$ we observe that there are only $d$ configurations of bonds in the statistical model that give a non-vanishing contribution. These configurations are the ones where each bond takes the same index $i$. Summing,
\begin{align}\label{ghzq}
    {\cal E}_n^{(\tq)} = \sum_{i=1}^d (|\lambda_i|^2)^{n^{\tq-1}}\equiv \sum_{i=1}^d a_i^n.
\end{align}
Here we have defined $0\leq a_i\equiv (|\lambda_i|^2)^{n^{\tq-2}} \leq 1$ for future use. 
When we identify party $A_\ta$ and $A_\tb$ as the same party, the state on the $\tq-1$ parties is isomorphic to the generalized GHZ state with the same coefficients $\lambda_i$. The lattice breaks up into  $n$ copies of $\tq-2$ dimensional lattice with the new $\tq-1$ partite generalized GHZ state as the Boltzmann weight. We get
\begin{align}\label{ghzq1}
    ({\cal E}_n^{(\tq-1)}|_{(A_\ta A_\tb)})^n= \Big(\sum_{i=1}^d (|\lambda_i|^2)^{n^{\tq-2}}\Big)^n= \Big(\sum_{i=1}^d a_i\Big)^n.
\end{align}
The equations \eqref{ghzq} and \eqref{ghzq1} have the same homogeneity in $a_i$. Let us rescale $a_i$ such that $\sum_i a_i=1$. As $0\leq a_i\leq 1$ even after rescaling, it is clear that the inequality $\scg$ is obeyed.

\section{Neighborhood of factorized states}\label{sec-factorized}
In this section, we will consider a $\tq$-partite pure state such that the density matrix after tracing out one of the parties is almost completely factorized. In other words, if we take the  traced-out party to be $A_\tq$, the density matrix takes the form
\begin{align}\label{almost-factorized}
    \rho_{A_\1,\ldots, A_{\tq-1}}=\rho_{A_1}\otimes \ldots \otimes \rho_{A_{\tq-1}} +\delta\rho_{A_\1,\ldots, A_{\tq-1}}.
\end{align}
Here we have taken the full density matrix $\rho_{A_\1,\ldots, A_{\tq-1}}$ and each factor $\rho_\ta$ in the leading order factorized density matrix to be trace normalized. So the trace of the perturbation $\delta \rho$ is zero.

\subsection{Inequality $\rcg$}
The inequality $\rcg$  is saturated for all factorized density matrices at leading order after identification of any two parties $A_\ta$ and $A_\tb$ as discussed in section \ref{factorized}. It is then a non-trivial question whether the inequality $\rcg$ continues to hold at sub-leading orders.  We will also investigate the saturation of $\rcg$ at sub-leading orders. For $\tq=3$, we will be able to completely characterize the subspace of deformations $\delta \rho$ that saturates the inequality $\rcg$. In doing so we will show that the inequality $\rcg$ holds for state \eqref{almost-factorized} to all orders of perturbation theory in $\delta\rho$.


\subsection{Tri-partite states}
For simplicity, let us first consider the case of tri-partite state $\tq=3$. 
The measure ${\cal E}_n^{(3)}$ is graphically represented as the partition function on an $n\times n$ lattice as described earlier near figure \ref{q3}.
At leading order, the rows and columns of the $2d$ lattice decouple. The partition function is given by,
\begin{align}
    {\cal E}^{(3)}_n=\Big({\rm Tr}\rho_{A_1}^n\Big)^n \, \Big({\rm Tr}\rho_{A_2}^n\Big)^n.
\end{align}
Here the first factor comes from the $n$ horizontal strands of the lattice and the second factor comes from the $n$ vertical strands. A single  strand is ``necklace'' of $n$ $\rho$'s contributing ${\rm Tr} \rho^n$.

The first correction is obtained by replacing $\rho_{A_\1}\otimes \rho_{A_\2}$ at one of the vertices by $\delta \rho_{A_\1,A_\2}$.
\begin{align}\label{1-correction}
    \delta{\cal E}^{(3)}_n= n^2 \,\Big({\rm Tr}\rho_{A_1}^n  \, {\rm Tr}\rho_{A_2}^n\Big)^{n-1} {\rm Tr}_{A_2} \Big(\rho_{A_\2}^{n-1} {\rm Tr}_{A_\1} \Big(\rho_{A_\1}^{n-1} \delta \rho_{A_\1,A_\2}\Big)\Big).
\end{align}
Here $ {\rm Tr}_{A_\ta}$ stands for the trace over only $A_\ta$ party and the combinatorial factor $n^2$ follows because there are $n^2$ vertices where the perturbation $\delta\rho$ can appear. There are $n-1$ horizontal as well as vertical strands of $n$ factorized density matrices $\rho_{A_\1}$ and $\rho_{A_\2}$ respectively that are completely decoupled from $\delta_\rho$. They contribute the factor $({\rm Tr}\rho_{A_1}^n  \, {\rm Tr}\rho_{A_2}^n)^{n-1}$. The last and the most non-trivial factor is contributed by one horizontal strand of $\rho_{A_\1}$'s and one vertical strand of $\rho_{A_\2}$'s intersecting at a single point. At the intersection point, the two strands are glued together by replacing the leading order $\rho_{A_\1}\otimes \rho_{A_\2}$ by  $\delta \rho_{A_\1,A_\2}$. 

The quantity $({\cal E}_n^{(2)})^n$ which we want to compare with ${\cal E}_n^{(3)}$ is also given by the same lattice but with vertical bonds oriented horizontally (and horizontal bonds staying where they are). The contribution of the strands that don't involve $\delta\rho$ is the same as in ${\cal E}_n^{(3)}$. In fact, the contribution of two strands that do have $\delta\rho$ is also the same as before. The only difference is that the two strands in question are not oriented horizontally and vertically but rather both of them are oriented horizontally. So the right-hand side of equation \eqref{1-correction} is precisely what we obtain if we compute the first correction to the $({\cal E}_n^{(2)})^n$ also.
This shows that the inequality $\rcg$ remains saturated at first order. In fact, for the conjecture to hold, $\rcg$ must remain saturated at first order because the inequality sign could have been reversed by changing the sign of $\delta \rho$.

At second order, we have two insertions of $\delta \rho$ on the lattice of ${\cal E}_n^{(\3)}$. This can happen in two inequivalent ways. Either the insertions are collinear along a row or column or they are not. When the insertions are not collinear, we simply get the square of the multiplicative correction factor i.e. the last factor on the right-hand side of equation \eqref{1-correction}, one corresponding to each insertion. The contribution to the second-order change due to these configurations is
\begin{align}
    \frac{n^2(n-1)^2}{2} \,\Big({\rm Tr}\rho_{A_1}^n  \, {\rm Tr}\rho_{A_2}^n\Big)^{n-2} \Big({\rm Tr}_{A_2} \Big(\rho_{A_\2}^{n-1} {\rm Tr}_{A_\1} \Big(\rho_{A_\1}^{n-1} \delta \rho_{A_\1,A_\2}\Big)\Big)\Big)^2.
\end{align}
As in the case of single insertions, the two insertions when they are not collinear,  contribute the same to the change in $({\cal E}_n^{(2)})^n$. So the sign of the inequality $\rcg$ is determined by the configurations of $\delta \rho$ that are collinear.  

Let us insert $\delta \rho$ at the vertex $A$ of the lattice. The other insertion then is taken to be at $B$ which is to the right of $A$ separated by, say $m$ number of $\rho$'s. Alternatively, it can be at $C$ which is above $A$ separated by $m$ number of $\rho$'s. Let this pair of configurations be labeled by $m$. To calculate the change at the second order, we need to sum over such pairs for $m=0,\ldots, n-2$. However, we will show that the inequality $\rcg$ holds separately for the contribution of configuration with a given value of $m$. 

\begin{figure}[t]
    \begin{center}
        \includegraphics[scale=0.25]{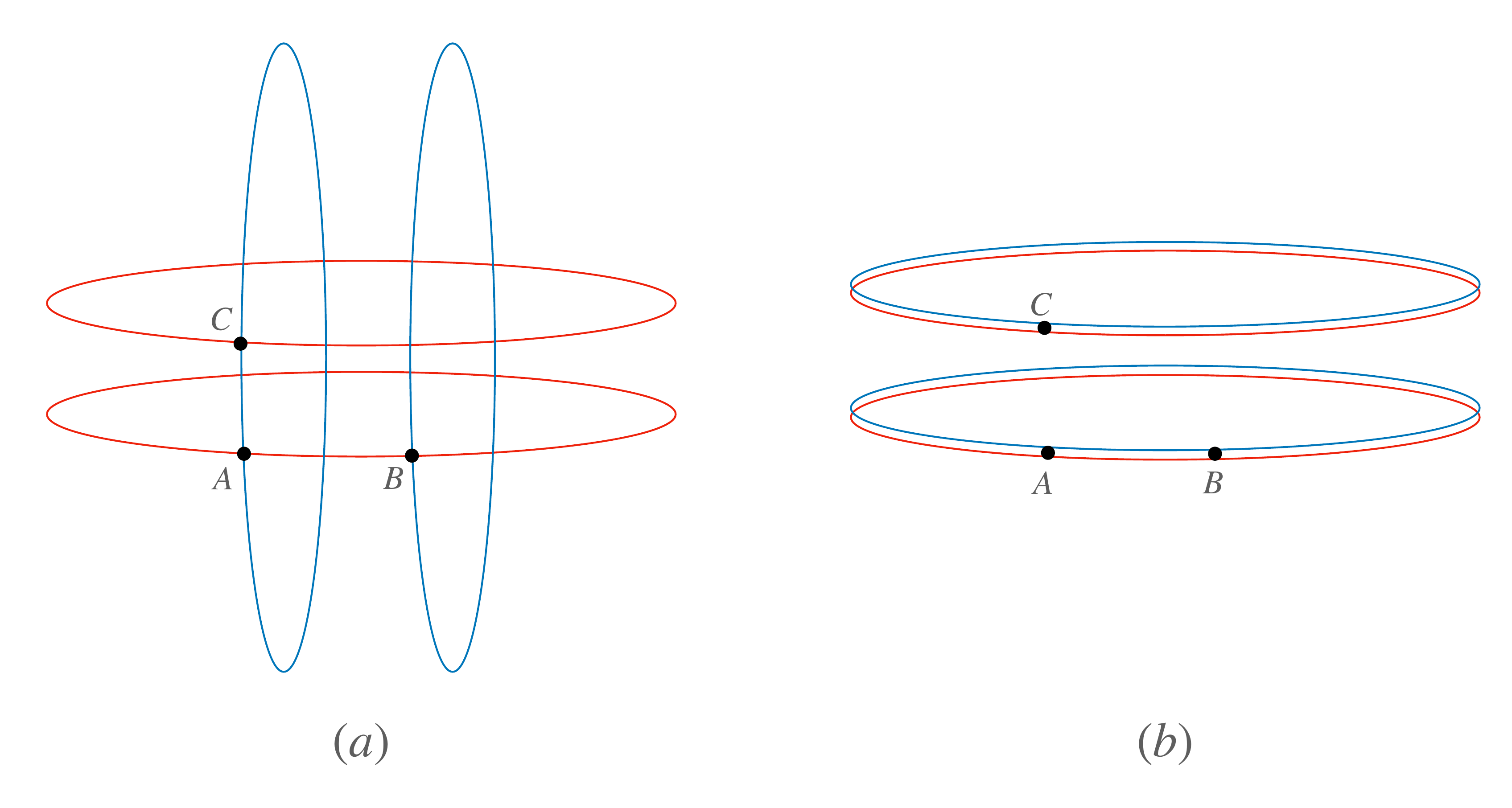}
    \end{center}
    \caption{The relevant part of the lattice that contributes to ${\cal E}_n^{(3)}$ and $({\cal E}_n^{(2)})^n$ respectively at second order. Without loss of generality, we fix one of the $\delta\rho$ insertions at $A$ in both diagrams. The other can be either at $B$ or $C$ separated by $m$ $\rho$'s.}\label{factorized}
\end{figure}
\begin{figure}[t]
    \begin{center}
        \includegraphics[scale=0.3]{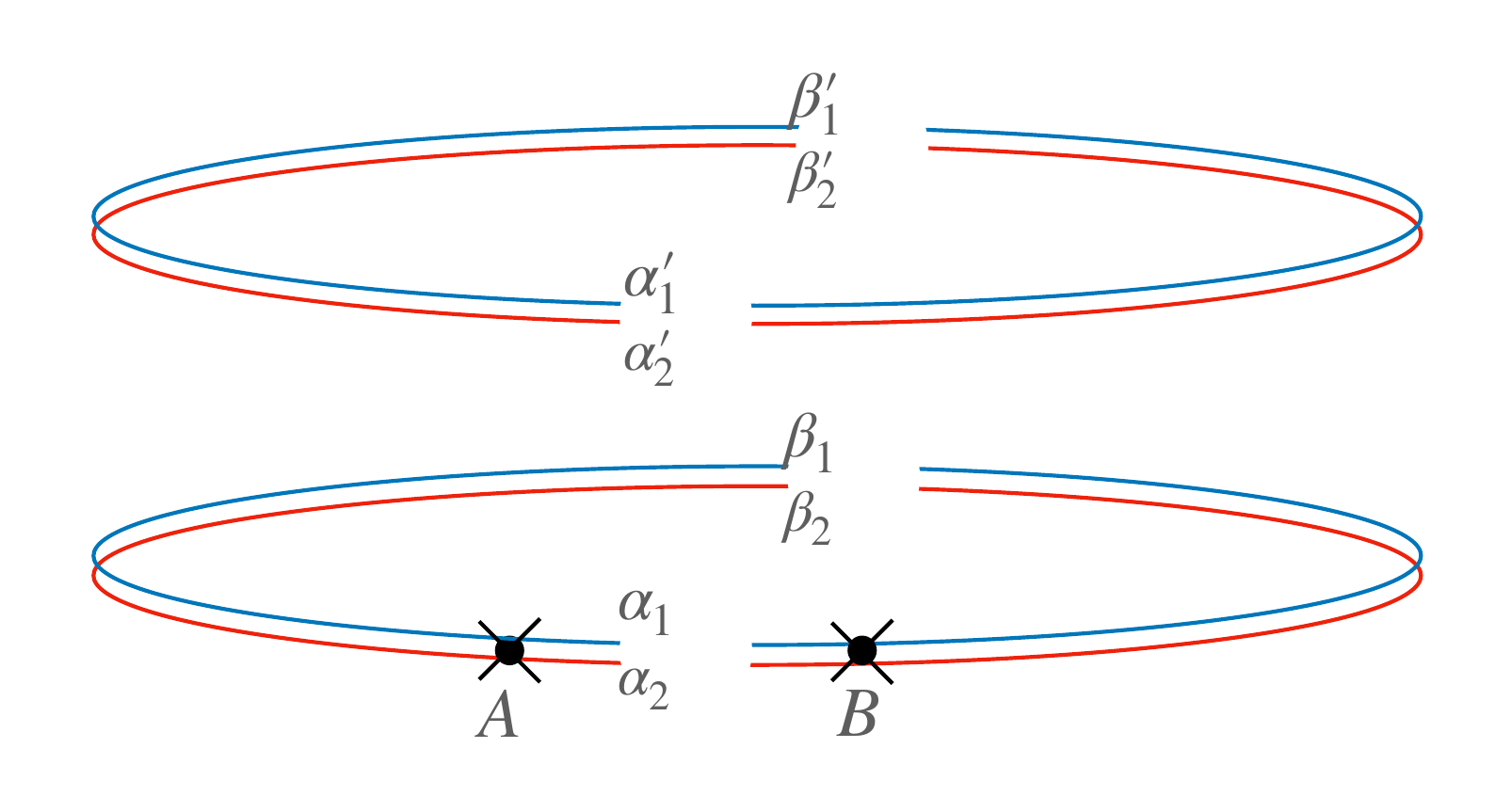}
    \end{center}
    \caption{Here we have taken the configuration where $\delta \rho$'s are inserted at $A$ and $B$ in the second diagram of figure \ref{factorized}. We have cut the four pairs of bonds precisely in the middle. This gives us a ket $|v_{ijkl}\rangle$ on the right and the corresponding bra on the left. Here the $i$ index corresponds to {\emph{both}} the ends of the topmost arc and so on. }\label{gluing-factorized}
\end{figure}

Let us first extract only the relevant part of the lattice pertaining to the configuration pair $m$. This is done in figure \ref{factorized}. The subfigure $(a)$  shows the part of the lattice relevant for computing the second order correction to ${\cal E}_n^{(3)}$ and the subfigure $(b)$ shows the same but for $({\cal E}_n^{(2)})^n$. The rest of the lattice only consists of leading order density matrices that are factorized and hence gives identical multiplicative contribution to both sides. The inequality is then determined only by the part of the lattice extracted in figure \ref{factorized}.
Consider the $\delta \rho$ insertions at $A$ and $B$ in subfigure $(b)$. Let us cut subfigure $(b)$ precisely in the middle as shown in figure \ref{gluing-factorized}. This gives us two pieces. One piece has a $\delta\rho$ insertion with $m/2$ and $(n-m)/2-1$ copies of $(\rho_{A_\1}\otimes \rho_{A_\2})$'s on either side. 
This gives the matrix 
\begin{align}\label{ml}
    (M_l)_{\alpha_1\alpha_2\alpha'_1\alpha'_2}^{\beta_1\beta_2\beta'_1\beta'_2} &= \Big((\rho_{A_\1}\otimes \rho_{A_\2})^{m/2} \cdot \delta\rho \cdot (\rho_{A_\1}\otimes \rho_{A_\2})^{(n-m)/2-1}\Big)_{\alpha_1\alpha_2}^{\beta_1\beta_2} \notag\\
    &\otimes
    \Big((\rho_{A_\1}\otimes \rho_{A_\2})^{n/2}\Big)_{\alpha'_1\alpha'_2}^{\beta'_1\beta'_2} 
\end{align}
on the left and a similar matrix $M_r$ on the right. Due to hermiticity of $\rho$ and $\delta\rho$, we have $M_r^\dagger =M_l$. When either $m$ or $n$ is odd, we get half-integer powers of $\rho$ in the expression \eqref{ml} for $M_l$. Using the fact that $\rho$ is not only hermitian but positive definite, we can define a hermitian matrix $\sqrt{\rho}$. Because of this, even in this case, we have the property $M_r^\dagger =M_l$.
Let us recall the standard matrix inner product $\langle M|N\rangle = {\rm Tr} M^\dagger N$. The contribution of the subfigure $(b)$ with $\delta\rho$'s  inserted at $A$ and $B$ can be thought of as the norm
\begin{align}
    {\rm Tr} (M_l \, M_r)=\langle M_l^\dagger |M_r\rangle = \langle M_r |M_r\rangle.
\end{align}
Consider the linear operators $P_i$  acting on $M_l$ and $M_r$ that swap the pair of indices $(\alpha_i,\beta_i)\leftrightarrow (\alpha'_i,\beta'_i)$. Because $P_i^2=1$, their eigenvalues are $\pm 1$. Now we are ready to prove the inequality. Consider
\begin{align}\label{positivemm}
    \langle M_l^\dagger| (1-P_1)(1-P_2) | M_r\rangle = \langle M_r |(1-P_1)(1-P_2)|M_r\rangle .
\end{align}

\begin{figure}[t]
    \begin{center}
        \includegraphics[scale=0.45]{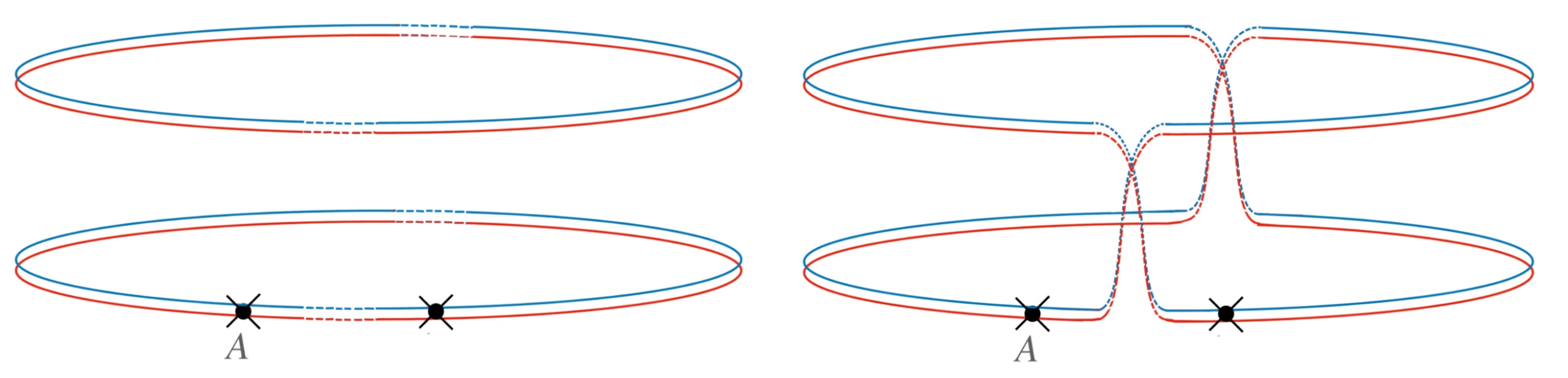}
    \end{center}
    \caption{$({\cal E}_n^{(2)})^n$ terms.}\label{glued1}
\end{figure}
\begin{figure}[t]
    \begin{center}
        \includegraphics[scale=0.45]{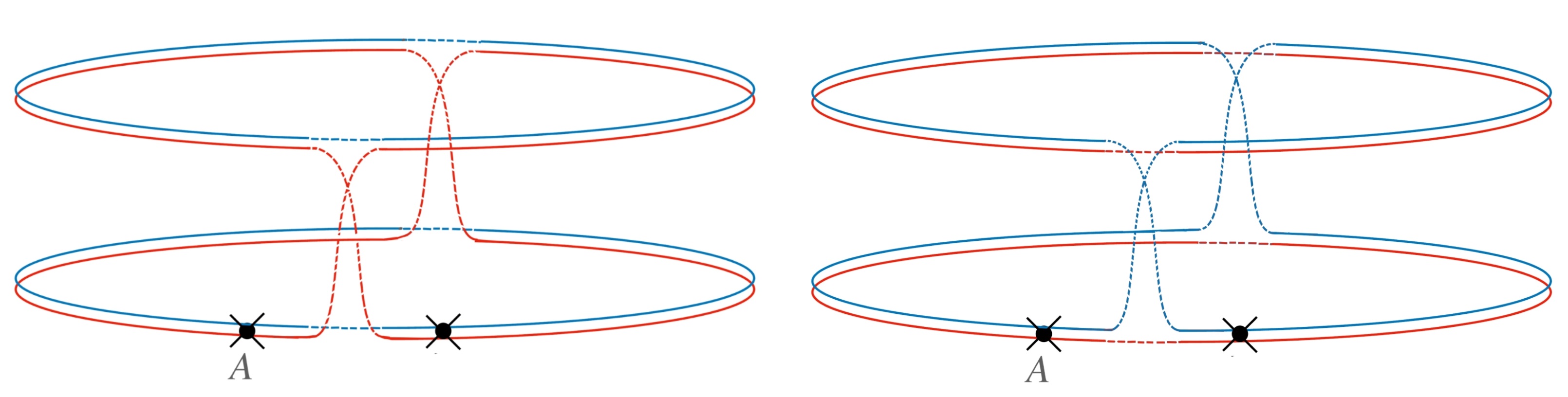}
    \end{center}
    \caption{${\cal E}_n^{(3)}$ terms.}\label{glued2}
\end{figure}
Using the fact that the eigenvalues of $P_i$ are $\pm 1$,  we conclude that the right-hand side is positive. To interpret the left-hand side, we observe, with the help of figure \ref{glued1} and figure \ref{glued2} that $\langle M_l^\dagger | 1+P_1P_2 | M_r\rangle$ is the contribution of the pair of configurations labeled by $m$ to $({\cal E}_n^{(2)})^n$ and $\langle M_l^\dagger| P_1+P_2 | M_r\rangle$ is the contribution of the pair of configurations labeled by $m$ to ${\cal E}_n^{(3)}$. Summing over $m$, we prove the inequality $\rcg$ for the tri-partite case.

Before moving to the general case of $\tq$ number of parties, let us analyze the saturation of the inequality $\rcg$ for the tri-partite state. Note that the operator $(1-P_1)/2$ squares to itself and so does the operator $(1-P_2)/\2$. Because $(1-P_1)(1-P_2)/4$ is a projector, the right hand side of \eqref{positivemm} is zero if and only if 
\begin{align}
    (1-P_1)(1-P_2) |M_r\rangle=0.
\end{align}
And as $P_1$ and $P_2$ commute, we write the solution as $M_r = M_r^{(1)}+M_r^{(2)}$ such that $(1-P_i)|M_r^{(i)} \rangle=0$. Let the perturbations associated to the solution $M_r^{(i)}$ be $\delta \rho^{(i)}$. Since $P_1 |M_{r}^{(1)}\rangle =|M_r^{(1)}\rangle$ and action of $P_1$ swaps the $(\alpha_1,\beta_1)$ index with $(\alpha'_1,\beta'_1)$, $|M_r^{(1)}\rangle$ must be symmetric with respect to such a swap. This means $\delta\rho_{A_1, A_2}^{(1)}$ must factorize as $\rho_{A_1}\otimes \delta\rho_{A_2}^{(1)}$. Similar arguments dictate that $\delta\rho_{A_1, A_2}^{(2)}= \delta\rho_{A_1}^{(2)} \otimes \rho_{A_2}$. All in all, for the inequality $\rcg$ to be saturated at second order, we need
\begin{align}
    \delta\rho_{A_1, A_2}= \rho_{A_1}\otimes \delta\rho_{A_2}+ \delta\rho_{A_1} \otimes \rho_{A_2}
\end{align}
for some arbitrary traceless hermitian matrices $\delta\rho_{A_1}$ and $\delta\rho_{A_2}$. Interestingly, the perturbed density matrix that saturates the inequality continues to be a factorized density matrix to linear order i.e.
\begin{align}
    \rho_{A_1}\rho_{A_2}+\delta\rho_{A_1,A_2} = (\rho_{A_1}+\delta\rho_{A_1})\otimes (\rho_{A_2}+\delta\rho_{A_2}) + {\cal O}(\delta \rho^2).
\end{align}
This means that we have to move along the tangent space of the factorized density matrix to continue to saturate the inequality. And because our starting point is an arbitrary factorized density matrix, it implies that in the neighborhood of the locus of factorized density matrices on two parties, the only density matrices that saturate the inequality $\rcg$ are the factorized density matrices. This leads to a natural conjecture that for tri-partite states, the inequality is saturated precisely for factorized density matrices or equivalently when the mutual information between the two parties in question vanishes.

\subsection{$\tq$-partite states}
The proof of the condition $\rcg$ for almost factorized $\tq$-partite state \eqref{almost-factorized} is very similar to that for the almost factorized tri-partite states. The inequality is saturated at leading and first order in the same way as before. In the analysis at second order, only the parties $A_\1$ and $A_\2$, that are being identified, enter the discussion. The figures \ref{factorized} and \ref{gluing-factorized} are drawn in the same way with parties $A_\1$ and $A_\2$ but with transverse $\tq-3$ dimensional ``strands'' corresponding to non-participating parties shooting off from $\delta\rho$ insertions at $A$ and $B$. 
In the tri-partite case, the hermiticity of $\delta\rho$ insertions at $A$ and $B$ played an important role in proving $\rcg$ at second order. For the $\tq$-partite case, the two party matrix  insertion at $A$ and $B$ is not $\delta\rho$ but rather $\widehat{\delta\rho}$ that is obtained by contracting $\delta\rho_{A_\1,A_\2,\ldots, A_{\tq-1}}$ to the  $\tq-3$ strands of $\rho_{A_\ta}$, $\ta=\3,\ldots, \tq-1$ of length $n$. Explicitly, this matrix $\widehat{\delta\rho}_{\alpha_\1\alpha_\2}^{\beta_\1\beta_\2}$ is 
\begin{align}
    \widehat{\delta\rho}_{\alpha_\1\alpha_\2}^{\beta_\1\beta_\2} = \delta\rho_{\alpha_\1\alpha_\2\ldots\alpha_{\tq-1}}^{\beta_\1\beta_\2 \ldots\beta_{\tq-1}} (\rho_{A_\3}^n)^{\alpha_\3}_{\beta_\3}\ldots (\rho_{A_{\tq-1}}^n)^{\alpha_{\tq-1}}_{\beta_{\tq-1}}.
\end{align}
Hermiticity of $\delta\rho$ and $\rho_{A_\ta}$'s implies the matrix insertion $\widehat{\delta\rho}$ is also hermitian. This provides the required step in the proof of $\rcg$ for almost factorized $\tq$-partite states. The saturation of $\rcg$ is analyzed in the same way. We get that the perturbation $\widehat{\delta\rho}$ must be a sum of two terms, one proportional to $\rho_{A_\1}$ and the other proportional to $\rho_{A_\2}$ for the inequality saturation i.e. 
\begin{align}\label{saturation}
    \widehat{\delta\rho} = \rho_{A_\1}\otimes \delta\rho _{A_\2} +\rho_{A_2}\otimes  \delta\rho_{A_\1}
\end{align}
for arbitrary small traceless hermitian matrices $\delta\rho_{A_\1}$ and $\delta\rho_{A_\2}$.

\subsubsection{Broadening the scope}
The above discussion suggests that we can broaden our analysis from the neighborhood of the completely factorized state to a more general state of the following form
\begin{align}\label{more-general}
    \rho_{A_\1,\ldots,A_{\tq-1}}= \rho_{A_\1}\otimes \rho_{A_\2} \otimes \rho^{\rm rest}_{A_\3,\ldots, A_{\tq-1}} +\delta\rho_{A_\1,\ldots,A_{\tq-1}}.
\end{align} 
Because the density matrix is factorized into party $A_\1$, party $A_\2$ and the rest, the proof of saturation of the inequality $\rcg$ at leading and first order goes through in the same way as before. At second order, again, the nontrivial problem is to show that the insertion matrix  $\widehat{\delta\rho}$ is hermitian. Instead of having decoupled transverse strands attached to the insertions like in the previous case, an entire $\tq-3$ dimensional torus is attached to it. This is the torus corresponding to the measure ${\cal E}_n^{(\tq-2)}(\rho^{\rm rest})$ with one of the $\rho^{\rm rest}$ vertex replaced by $\delta\rho$. Concretely, consider the matrix $H_n^{(\tq-2)}(\rho^{\rm rest})$ obtained by removing a vertex in the lattice corresponding to ${\cal E}_n^{(\tq-2)}(\rho^{\rm rest})$. Then,
\begin{align}
    \widehat{\delta\rho}_{\alpha_\1\alpha_\2}^{\beta_\1\beta_\2} = \delta\rho_{\alpha_\1\alpha_\2\ldots\alpha_{\tq-1}}^{\beta_\1\beta_\2 \ldots\beta_{\tq-1}} \Big(H_n^{(\tq-3)}(\rho^{\rm rest})\Big)^{\alpha_\3\ldots\alpha_{\tq-1}}_{\beta_\3 \ldots\beta_{\tq-1}}.
\end{align}
Because $H_n^{(\tq)}(\rho)$ is a hermitian matrix for general $\rho$, $\widehat{\delta\rho}$ is also hermitian. This shows that even for the more general state of the type \eqref{more-general}, the inequality $\rcg$ holds. The same argument as before shows that it is saturated if and only if $\widehat{\delta\rho}$ is the sum of two terms, one proportional to $\rho_{A_\1}$ and the other proportional to $\rho_{A_\2}$ as in equation \eqref{saturation}. 
What does this condition mean for $\delta \rho$ in equation \eqref{more-general}? We can't solve it generally for $\delta\rho$. However, we see that the following is certainly a solution to the condition \eqref{saturation}.
\begin{align}\label{special-sol}
    \delta\rho_{A_\1,\ldots,A_{\tq-1}} = \rho_{A_\1} \otimes \delta\rho^{(1)}_{A_\2,A_\3,\ldots,A_{\tq-1}}+ \rho_{A_\2} \otimes \delta\rho^{(2)}_{A_\1, A_\3, \ldots,A_{\tq-1}}.
\end{align}
Interestingly, in this space of solutions, the mutual information between parties $A_\1$ and $A_\2$ vanishes at first order in perturbation theory. This is seen as follows. Let us construct $\rho_{A_\1,A_\2}$ from the full $\rho$ given in \eqref{more-general} with $\delta\rho$ given by the special solution \eqref{special-sol}.
\begin{align}
    \rho_{A_\1,A_\2}&=\rho_{A_\1}\otimes \rho_{A_\2}+ \rho_{A_\1} \otimes \delta \rho_{A_2} +\delta\rho_{A_1}\otimes \rho_{A_2}\\
    \delta \rho_{A_2} &\equiv {\rm Tr}_{A_\3,\ldots,A_{\tq-1}} \delta\rho^{(1)}_{A_\2,A_\3,\ldots,A_{\tq-1}} \notag\\
    \delta \rho_{A_1} &\equiv {\rm Tr}_{A_\3,\ldots,A_{\tq-1}} \delta\rho^{(2)}_{A_\2,A_\3,\ldots,A_{\tq-1}}. \notag
\end{align}
Up to corrections of ${\cal O}(\delta\rho^2)$, the right-hand side can be written as a factorized density matrix $(\rho_{A_\1}+\delta\rho_{A_\1})\otimes (\rho_{A_\2}+\delta\rho_{A_\2})$ showing that the mutual information between $A_\1$ and $A_\2$ vanishes at first order. However, in general, the condition of saturation of inequality $\rcg$ and the vanishing of mutual information seem unrelated.

\subsection{Multi-entropy for almost factorized states}
In \cite{Gadde:2022cqi}, authors defined Multi-entropy from Renyi multi-entropy as the limit
\begin{align}
    S^{(\tq)}\equiv \lim_{n\to 1} \,S_n^{(\tq)}= 
    \partial_n{\cal E}_n^{(\tq)}|_{n=1}.
\end{align}
In the second equality, we have used the fact that ${\cal E}_1^{(\tq)}=1$. 
The definition of multi-entropy requires analytically continuing $S_n^{(\tq)}$, which is defined for integer $n$, to complex values of $n$. The precise nature of this analytic continuation is a delicate question. We will not deal with it here. In this paper, we will perform the analytic continuation that is most straightforward. We will first expand ${\cal E}_n^{(\tq)}(\rho)$ in powers of $\delta\rho$ and analytically continue each term in the perturbative expansion. Finally, we will show that the inequality $\rcg$ holds for the multi-entropy up to ${\cal O}(\delta\rho^2)$ just like for integer $n$'s.

\subsubsection*{Leading order}
At the leading order, we have already computed the Renyi multi-entropy in \eqref{m-renyi-factorized}. Using the fact that the analytic continuation of the Renyi entropy $S^{(2)}_n(\rho)$ to $n=1$ gives Von Neumann entropy $S(\rho)$, we get
\begin{align}\label{leading-multi}
    S^{(\tq)}(\rho)= \sum_{\ta=1}^{\tq-1} S(\rho_{A_\ta}) + {\cal O}(\delta\rho).
\end{align}
To compute the $n\to 1$ limit at higher orders, it is useful to diagonalize the leading order factorized density matrix, let's call it $\rho_0$, within each party. Let $v_{\alpha_\ta}$ be the eigenvectors of $\rho_{A_\ta}$  with eigenvalues $\lambda_{\alpha_\ta}$. Then
\begin{align}
    {\cal E}_n^{(\tq)}(\rho)|_{\rm leading}=\prod_{\ta=1}^{\tq-1}\Big(\sum_{\alpha_\ta=1}^{d_\ta} \lambda_{\alpha_\ta}^n\Big)^{n^{\tq-2}}\equiv (A_n)^{n^{\tq-2}}.
\end{align}
The second equality serves as a definition of $A_n$. This expression can be straightforwardly analytically continued in $n$. It has the property that $A_1=1$ and $A'_1=\sum_{\ta} \sum_{\alpha_\ta=1}^{d_\ta}\lambda_{\alpha_\ta} \log(\lambda_{\alpha_\ta})$. This gives us \eqref{leading-multi} as expected.

\subsubsection*{First order}
The first order correction to ${\cal E}_n^{(\tq)}$ is obtained by replacing one vertex of the lattice of $\rho_0$'s with $\delta\rho$. This gives
\begin{align}
    {\cal E}_n^{(\tq)}|_{1\,{\rm st}}= {\rm Tr}\, H_n^{(\tq)}(\rho_0) \, \delta\rho.
\end{align} 
This lattice consists of decoupled $1$-dimensional strands except for the strands that are attached to the $\delta\rho$ vertex. We get
\begin{align}
    {\cal E}_n^{(\tq)}|_{1\,{\rm st}}&= n^{\tq-1}(A_n)^{n^{\tq-2}-1} B_n\\
    B_n&\equiv \sum_{\alpha_\1=1}^{d_\1}\ldots \sum_{\alpha_{\tq-1}=1}^{d_{\tq-1}} (\lambda_{\alpha_\1}^{n-1} \ldots \lambda_{\alpha_{\tq-1}}^{n-1}) \delta\rho_{\alpha_\1\ldots \alpha_{\tq-1}}^{\alpha_\1\ldots \alpha_{\tq-1}}.\notag
\end{align}
Here $B_n$ can also be analytically continued by promoting the exponent of $\lambda$ to a complex number. We have $B_1=0$, due to the traceless-ness of $\delta\rho$. So $\partial_n {\cal E}_n^{(\tq)}|_{1\,{\rm st}}$ at $n=1$ gets nonzero contribution when the derivative hits $B_n$. We get $(\partial_n {\cal E}_n^{(\tq)}|_{1\,{\rm st}})|_{n=1} =B'_1$. Hence
\begin{align}\label{order1}
    S^{(\tq)}(\rho)|_{1\,{\rm st}}= -B_1'=-\sum_{\ta=\1}^{\tq-1} \sum_{\alpha_\ta=1}^{d_\ta}\left(
         \log \lambda_{\alpha_{\ta}} \,(\delta\rho_{A_\ta})_{\alpha_\ta}^{\alpha_{\ta}}\right)=-\sum_{\ta=\1}^{\tq-1} {\rm Tr}(\delta\rho_{A_\ta} \log \rho_{A_\ta}).
\end{align} 
Here $\delta\rho_{A_\ta}$ is obtained from $\delta\rho$ by tracing over all the parties except $A_\ta$. Using the fact that $-\log \rho_{A_\ta}$ is the modular Hamiltonian $K_\ta$, interestingly the first correction to the multi-entropy is given by $\sum_{\ta=1}^{\tq-1} \langle K_\ta\rangle_{\delta\rho_{A_\ta}}$ as it would be for the sum of entanglement entropies.

\subsubsection*{Second order}
The computation of ${\cal E}_n^{(\tq)}$ is a little more involved. It comes from two types of configurations. One where the two insertions of $\delta\rho$ are non-collinear and the other where they are collinear. The contribution of the non-collinear configurations
\begin{align}
    {\cal E}_n^{(\tq)}|_{2\,{\rm nd, non-coll}}= \frac12 n^{\tq-1}(n^{\tq-1}-(n-1)(\tq-1)-1)(A_n)^{n^{\tq-2}-2} B_n^2.
\end{align}
The first factor is a combinatorial one that imposes for the second insertion that it is not collinear to the first one. It is easy to see that the value and the derivative at $n=1$ of ${\cal E}_n^{(\tq)}|_{2\,{\rm nd, non-coll}}$  is zero. Here we use $B_1=0$. We get the non-zero contribution at second order from configurations where the two insertions of $\delta\rho$ are collinear. Without loss of generality, let us take them to be separated in $A_\1$ directions with $m$ insertions on $\rho_{A_\1}$ between them. The entire lattice is decoupled into $1$-dimensional lattices, except for the sub-lattice that is attached to the two $\delta\rho$ insertions.  For all parties, except for $A_\1$, two of the $n^{\tq-2}$ strands are attached to $\delta\rho$ and for party $A_\1$, one of the strands is attached to $\delta\rho$'s. The contribution from the strands that are not attached to $\delta\rho$'s is then
\begin{align}
    \Big(\sum_{\alpha_\1=1}^{d_\1}\lambda_{\alpha_\1}^n\Big)\prod_{\ta=1}^{\tq-1} \Big(\sum_{\alpha_\ta=1}^{d_\ta}\lambda_{\alpha_\ta}^n\Big)^{n^{\tq-2}-2}=\Big(\sum_{\alpha_\1=1}^{d_\1}\lambda_{\alpha_\1}^n\Big) A_n^{n^{\tq-2}-2}\equiv J_n^{(1)}.
\end{align}
The contribution from the strands that are attached to $\delta\rho$ is
\begin{align}
    L_{n}^{(1)}& \equiv (B^{(1)}_n)_{\alpha_\1}^{\beta_\1} (B^{(1)}_n)^{\alpha_\1}_{\beta_\1} \sum_{m=0}^{n-2}\lambda_{\alpha_\1}^m\lambda_{\beta_1}^{n-m-2} = (B^{(1)}_n)_{\alpha_\1}^{\beta_\1} (B^{(1)}_n)^{\alpha_\1}_{\beta_\1} \frac{\lambda_{\alpha_{\1}}^{n-1}-\lambda_{\beta_{\1}}^{n-1}}{\lambda_{\alpha_{\1}}-\lambda_{\beta_{\1}}} .\notag \\
    (B^{(1)}_n)_{\alpha_\1}^{\beta_\1}&\equiv \sum_{\alpha_\2=1}^{d_\2}\ldots \sum_{\alpha_{\tq-1}=1}^{d_{\tq-1}} (\lambda_{\alpha_\2}^{n-1} \ldots \lambda_{\alpha_{\tq-1}}^{n-1}) \delta\rho_{\alpha_\1\ldots \alpha_{\tq-1}}^{\beta_\1\ldots \alpha_{\tq-1}}
\end{align}
In the first line, we have summed over different values of $m$. Summing over the party along which the $\delta\rho$ insertions are separated (which was taken to be $A_\1$ in the above discussion), we get
\begin{align}\label{2nd-correction}
    {\cal E}_n^{(\tq)}|_{2\,{\rm nd, coll}}= \frac{n^{\tq-1}}{2}\sum_{\ta=1}^{\tq-1}  J^{(\ta)}_{n} L^{(\ta)}_{n}.
\end{align}
Here the first factor is a combinatorial factor associated with the symmetry of this configuration. The final answer \eqref{2nd-correction} admits a straightforward analytic continuation in $n$. The derivative of ${\cal E}_n^{(\tq)}|_{2\,{\rm nd, coll}}$ at $n=1$ contributes only when the derivative hits the ratio appearing in $L_n^{(\ta)}$. All the other values of $n$ must be set to $1$. We get the matrix $B_1^{\ta}=\delta\rho_{A_\ta}$ and also $J_1^{(1)}=1$. Hence,
\begin{align}\label{order2}
    S^{(\tq)}(\rho)|_{2\,{\rm nd}}= -\frac12 \sum_{\ta=1}^{\tq-1} {\rm Tr}(\delta\rho_{A_\ta}^2 M_\ta),\qquad (M_{A_\ta})_{\alpha_\ta}^{\beta_\ta}\equiv\frac{\log \lambda_{\alpha_{\ta}}-\log \lambda_{\beta_{\ta}}}{\lambda_{\alpha_{\ta}}-\lambda_{\beta_{\ta}}} 
\end{align}
The symmetric and real matrix $M_{A_\ta}$ is defined by specifying its components in the distinguished basis (the basis of eigenvectors of $\rho_{A_\ta}$) $v_{\alpha_\ta}$ as shown above.
Combining the results \eqref{leading-multi}, \eqref{order1} and \eqref{order2}, we get
\begin{align}\label{multi-full}
    S^{(\tq)}(\rho)=\sum_{\ta=1}^{\tq-1}\left( S(\rho_{A_\ta})+ \langle K_{\ta}\rangle_{\delta\rho_{A_\ta}} -\frac12 {\rm Tr}(\delta\rho_{A_\ta}^2 M_{A_\ta}) \right)+{\cal O}(\delta\rho^3).
\end{align}

\subsection{Monotonicity for multi-entropy}
In this section, we will show that the multi-entropy \eqref{multi-full} obeys the inequality  $\rcg$ in the neighborhood of the factorized state. At leading order, from equation \eqref{leading-multi}, it is easy to see that the inequality is saturated. In fact, it is saturated even for the first correction as in the case of integer $n$. This is seen as follows. We use $\log (\rho_{A}\otimes \rho_B)=\log(\rho_A) \otimes {\mathbb I}+{\mathbb I}\otimes \log(\rho_{B})$. We need only focus on the terms contributed by the two parties, say $A_\1$ and $A_\2$, that are being identified. The other terms are the same on both sides of $\rcg$. For the relevant parties we have,
\begin{align}
    -{\rm Tr}(\delta\rho_{A_\1,A_\2} \log  \rho_{A_\1,A_\2})= -{\rm Tr}(\delta\rho_{A_\1} \log  \rho_{A_\1})-{\rm Tr}(\delta\rho_{A_\2} \log  \rho_{A_\2}).
\end{align}
This shows that $\rcg$ continues to be saturated at first order. 

Before we look at the multi-entropy at second order. 
Let us define a symmetric matrix $(N_A)_\alpha^\beta= \sqrt{(M_{A})_\alpha^\beta}$. Just like $M$ we are defining the matrix $N$ by specifying its entries in a chosen basis. Because, $(M_{A})_\alpha^\beta$ are positive, the entries of $N$ are  real. Hence $N$ is hermitian. Let us also note a useful property of matrix $M$. 
\begin{align}
    (M_{A_\1, A_\2})_{\alpha_\1\alpha_\2}^{\beta_\1\beta_\2}=\frac{\log (\lambda_{\alpha_{\1}} \lambda_{\alpha_{\2}})-\log (\lambda_{\beta_{\1}} \lambda_{\beta_{\2}})}{\lambda_{\alpha_{\1}}\lambda_{\alpha_{\2}}-\lambda_{\beta_{\1}}\lambda_{\beta_{\2}}}. 
\end{align}
So,
\begin{align}\label{M-property}
    (M_{A_\1, A_\2})_{\alpha_\1\alpha_\2}^{\alpha_\1\beta_\2}=\lambda_{\alpha_\1}^{-1} M_{A_\2},\qquad (M_{A_\1, A_\2})_{\alpha_\1\alpha_\2}^{\beta_\1\alpha_\2}=\lambda_{\alpha_\2}^{-1} M_{A_\1}.
\end{align}

Motivated by our proof of the inequality $\rcg$ for integer $n$,
Consider the two matrices,
\begin{align}
    (M_l)_{\alpha_1\alpha_2\alpha'_1\alpha'_2}^{\beta_1\beta_2\beta'_1\beta'_2} &= (\delta\rho_{A_\1, A_\2})_{\alpha_\1\alpha_\2}^{\beta_\1\beta_\2} (N_{A_\1, A_\2})_{\alpha_\1\alpha_\2}^{\beta_\1\beta_\2} \,
    \Big((\rho_{A_\1}\otimes \rho_{A_\2})^{1/2}\Big)_{\alpha'_1\alpha'_2}^{\beta'_1\beta'_2} \notag\\
    &= \lambda_{\alpha'_1} \lambda_{\alpha'_2} (\delta\rho_{A_\1, A_\2})_{\alpha_\1\alpha_\2}^{\beta_\1\beta_\2} (N_{A_\1, A_\2})_{\alpha_\1\alpha_\2}^{\beta_\1\beta_\2} \delta_{\alpha'_1}^{\beta'_1}\delta_{\alpha'_2}^{\beta'_2}.
\end{align}
This is a hermitian matrix. In the second equality, we have used the fact that $\alpha_\ta$ and $\beta_\ta$ are eigenvectors $\rho_{A_\ta}$ and hence diagonalize it. To match with the notation from before, let us also define $M_r\equiv M_l$. We are now ready to prove the inequality at second order. As before, consider
\begin{align}\label{ineq-2}
    \langle M_l^\dagger | (1-P_1)(1-P_2)|M_r\rangle = \langle M_r | (1-P_1)(1-P_2)|M_r\rangle.
\end{align}
Here we have used $M_l=M_r$ and $M_r^\dagger = M_r$. The operators $P_1$ and $P_2$ act in the same way on indices $(\alpha,\alpha',\beta,\beta')$ as defined near equation \eqref{positivemm}. Because the eigenvalues of $P_i$ are $\pm 1$, the quantity \eqref{ineq-2} is positive. Now we simply note that $\langle M_l^\dagger | 1+P_1 P_2 |M_r\rangle$ gives $-2S^{(\tq)}(\rho_{A_\1,A_\2})$ and $\langle M_l^\dagger | P_1 + P_2 |M_r\rangle$ gives $-2S^{(\tq)}(\rho_{A_\1})-2S^{(\tq)}(\rho_{A_\2})$. In showing the latter, we need to use the property \eqref{M-property}.

As before, for the inequality $\rcg$ to be saturated we need $\delta\rho_{A_\1, A_\2}$ to be sum of two terms, one proportional to $\rho_{A_\1}$ and the other proportional to $\rho_{A_\2}$ i.e. we need, 
\begin{align}
    \delta\rho_{A_\1, A_\2}=\rho_{A_\1}\delta\rho_{A_\2}+\rho_{A_\2}\delta\rho_{A_\1}.
\end{align}
Recall, $\delta\rho_{A_\ta}$ is obtained from the full $\delta\rho_{A_\1,\ldots, A_{\tq-1}}$ by tracing out all the parties except $A_\ta$. We can't solve this condition in full generality for $\delta\rho_{A_\1,\ldots, A_{\tq-1}}$ but we can observe that 
\begin{align}
    \delta\rho_{A_\1,\ldots,A_{\tq-1}} = \rho_{A_\1} \otimes \delta\rho^{(1)}_{A_\2,A_\3,\ldots,A_{\tq-1}}+ \rho_{A_\2} \otimes \delta\rho^{(2)}_{A_\1, A_\3, \ldots,A_{\tq-1}}.
\end{align}
is certainly a solution. As argued near equation \eqref{special-sol}, this solution leads to a density matrix that has vanishing mutual information to first order in $\delta\rho$. But more generally, the saturation of the inequality $\rcg$ for multi-entropy and the vanishing of mutual information seem unrelated. 

\subsection{Inequality $\rpt$}
As discussed in section \ref{factorized}, the inequality $\rpt$ after partially tracing over say party $A_{\1}$ of the factorized density matrix is strictly obeyed except when the $n$-th Renyi entropy of $\rho_{A_\1}$ vanishes. In this case, the inequality is $\rpt$ is saturated. So precisely in this case, we can investigate $\rpt$ at the next order. 

At leading order the $\tq-1$ dimensional lattice computing ${\cal E}_n^{(\tq-1)}$ breaks up into $n^{\tq-2}$ number of one-dimensional lattices, one set for each party. This gives
\begin{align}
    {\cal E}_n^{(\tq)}= \Big({\rm Tr}\rho_{A_1}^n\Big)^{n^{\tq-2}}\ldots \Big({\rm Tr}\rho_{A_\tq}^n\Big)^{n^{\tq-2}}.
\end{align}
The first correction is obtained by replacing $\rho_{A_{\1}}\otimes \ldots \otimes \rho_{A_\tq}$ at of the vertices of the lattice by the perturbation $\delta \rho_{A_\1,\ldots,A_\tq}$. The correction is given by 
\begin{align}
    {\cal E}_n^{(\tq)}= n^{\tq-1}\Big({\rm Tr}\rho_{A_1}^n\Big)^{n^{\tq-2}-1}\ldots \Big({\rm Tr}\rho_{A_\tq}^n\Big)^{n^{\tq-2}-1} {\rm Tr}_{A_\1,\ldots,A_\tq}\Big((\rho_{A_\1}\otimes \ldots \otimes \rho_{A_\tq})^{n-1}\cdot \delta \rho_{A_\1,\ldots,A_\tq} \Big).\notag
\end{align}
Here the first factor $n^{\tq-1}$ is a combinatorial factor corresponding to the number of possible lattice points where the replacement can occur. The next factor is the contribution of $n^{\tq-2}-1$ number of one-dimensional lattices whose contribution remains unaffected. The last factor is the most interesting one. It corresponds to the $\tq-1$ strands of the lattice that are connected to $\delta\rho$ vertex. We will rewrite the last term as
\begin{align}
    &{\rm Tr}_{A_\1,\ldots,A_\tq}\Big((\rho_{A_\1}\otimes \ldots \otimes \rho_{A_\tq})^{n-1}\cdot \delta \rho_{A_\1,\ldots,A_\tq} \Big)= {\rm Tr}_{A_\1} \,\,\rho_{A_\1}^{n-1} \cdot \widetilde{\delta\rho}_{A_\1},\\
    {\rm where}\quad &\widetilde{\delta\rho}_{A_\1} \equiv  {\rm Tr}_{A_\2,\ldots,A_\tq}\Big((\rho_{A_\1}\otimes \ldots \otimes \rho_{A_\tq})^{n-1}\cdot \delta \rho_{A_\1,\ldots,A_\tq} \Big).\notag
\end{align}
Let us label the normalized eigenvectors of the combined density matrix $\rho_{A_\1}\otimes \ldots \otimes \rho_{A_\tq}$ by $|v_\beta\rangle $ and the corresponding eigenvalues by $\lambda_\beta$. Then the explicit form of $\widetilde{\delta\rho}_{A_\1}$ can be written as
\begin{align}
    \widetilde{\delta\rho}_{A_\1}  = \sum_\beta \lambda_\beta^{n-1} \langle v_\beta| \delta \rho_{A_\1,\ldots,A_\tq}|v_\beta\rangle.
\end{align}


Now let us compute $({\cal E}_n^{(\tq-1)})^n$ for the density matrix obtained tracing over party $A_\1$. The only difference is that instead of the last factor ${\rm Tr}_{A_\1} \,\rho_{A_\1}^{n-1} \cdot \widetilde{\delta\rho}_{A_\1}$, we get the product of traces $({\rm Tr} \,\rho_{A_\1})^{n-1} ({\rm Tr} \,\widetilde{\delta\rho}_{A_\1})= {\rm Tr}\, \widetilde{\delta\rho}_{A_\1}$ here we have replaced ${\rm Tr} \rho_{A_\1}=1$. So comparison of ${\cal E}_n^{(\tq)}$ and $({\cal E}_n^{(\tq-1)})^n$ amounts to the comparison between 
\begin{align}
    {\rm Tr}_{A_\1} \,\,\rho_{A_\1}^{n-1} \cdot \widetilde{\delta\rho}_{A_\1} \qquad {\rm and}\qquad {\rm Tr}\, \widetilde{\delta\rho}_{A_\1}.
\end{align}
Recall that we want to compare the quantities at the first order only for the case when $n$-th Renyi entropy of $\rho_{A_\1}$ vanishes. This condition implies that the rank of $\rho_{A_\1}$ is $1$. Let us choose an arbitrary eigenbasis of $\rho_{A_\1}$ as $|u_\alpha\rangle$ such that the eigenvalue for $\alpha=1$ is $1$ and for $\alpha\neq 1$ it is $0$. 
This means that $(\rho_{A_\1})_{11}=1$ and all other components are $0$. For this case, ${\rm Tr}_{A_\1} \,\,\rho_{A_\1}^{n-1} \cdot \widetilde{\delta\rho}_{A_\1} = (\widetilde{\delta\rho}_{A_\1})_{11}$.
We will show ${\rm Tr}\, \widetilde{\delta\rho}_{A_\1} \geq (\widetilde{\delta\rho}_{A_\1})_{11}$ as follows. First notice that 
\begin{align}\label{drpositive}
    \langle u_\alpha| \langle v_\beta| \rho_{A_\1}\otimes\rho_{A_\2} \ldots\otimes \rho_{A_\tq}+\delta\rho |u_\alpha\rangle| v_\beta\rangle \geq 0, \qquad {\rm for}\quad \alpha\neq 1,
\end{align}
thanks to the positivity of the perturbed density matrix. Now, the expectation value of the leading order factorized density matrix $\rho_{A_\1}\otimes\rho_{A_\2} \ldots\otimes \rho_{A_\tq}$ vanishes for $\alpha\neq 1$ thanks to the property $(\rho_{A_\1})_{\alpha\alpha}=0$ for $\alpha\neq 1$. This provides us the inequality $\langle u_\alpha| \langle v_\beta|\delta\rho |u_\alpha\rangle| v_\beta\rangle \geq 0$ for $\alpha\neq 1$. Summing this with positive weight $\lambda_\beta^{n-1}$, we get 
\begin{align}
    (\widetilde{\delta\rho}_{A_\1})_{\alpha\alpha} \geq 0 \qquad {\rm for} \quad \alpha\neq 1.
\end{align}
Because ${\rm Tr}\, \widetilde{\delta\rho}_{A_\1}- (\widetilde{\delta\rho}_{A_\1})_{11}$ simply consists of the sum above diagonal entries, we prove the inequality that we set out to prove. 

We now ask, what is the space of perturbations that continues to saturate the inequality $\rpt$ also at first order? This would happen if and only if  $(\widetilde{\delta\rho}_{A_\1})_{\alpha\alpha}=0$ for all $\alpha\neq 1$. We can not solve this condition for $\delta\rho$ in general but it is easy to check that $\delta\rho$ of the following form obeys this condition.
\begin{align}
    \delta\rho=\delta\rho_{A_\1}\otimes \delta\rho_{A_\2,\ldots,A_\tq},
\end{align}
such that the only element of $\delta\rho_{A_\1}$ that is non-zero is $(\delta\rho_{A_\1})_{11}$.

Unlike the case of inequality $\rcg$, the inequality $\rpt$ did not vanish at first order in $\delta\rho$. In the case of $\rpt$, we could not have changed the sign of $\delta\rho$ to flip the sign of the inequality because, the leading order, $\rho_{A_\1}$ that we started with had rank $1$ and lived on the boundary of parameter space. Changing the sign of $\delta\rho$ would take us outside the domain of valid density matrices.

\section{Probabilities}\label{probs}
The probability distribution for $\tq-1$ events can be thought of as a completely diagonal density matrix on $\tq-1$ parties. Equivalently, the probability distributions can be thought of as classical states. In this section, we will study the inequalities $\rcg$ and $\rpt$ for such states. 

Before we consider such classical density matrices let us consider a more general $\rho_{A_\1,\ldots, A_\tq}$ with the following property that the ``transfer matrix'' $T$ in the direction $\ta$ constructed out $\rho$'s is hermitian and positive definite. This transfer matrix is obtained by constructing a $\tq-2$ dimensional lattice of $\rho$ vertices that is orthogonal to $\ta$ direction. This lattice has $n^{\tq-2}$ fundamental and anti-fundamental indices of party $A_\ta$. In other words, it is a matrix acting on ${\cal H}_{A_\ta}^{\otimes n^{\tq-2}}$. If $T$ is hermitian and positive definite then it can be thought of as a density matrix itself (not normalized to $1$). The inequality $\rpt$ is then understood as the inequality between the ordinary Renyi-entropy and the $n$-th power of the norm. The normalization of $T$ does not matter because both sides of $\rpt$ have the same homogeneity in $T$ viz. $n$ so we might as well normalize $T$. Then $\rpt$ is the statement that the Renyi entropy is less than $1$.
Now, because the classical density matrix is completely diagonal with positive entries, the transfer matrix $T$ constructed from it is also diagonal with positive entries. In particular, $T$ is hermitian and positive definite. This proves the inequality $\rpt$ for classical density matrices.

Now we move our attention to proving $\rcg$. This turns out to be more involved. Motivated by the above argument about the transfer matrices. We can again reduce the $\tq-1$ partite problem only to the two parties $A_\ta$ and $A_\tb$ in question. This is done by constructing a $\tq-3$ dimensional lattice that is orthogonal to directions $\ta$ and $\tb$, the resulting object is a matrix acting on ${\cal H}_{A_\ta}^{\otimes n^{\tq-3}}\otimes {\cal H}_{A_\ta}^{\otimes n^{\tq-2}}$. Thanks to the positive and diagonal nature of the classical density matrix, this new ``codimension-2'' transfer matrix can also be thought of as a classical density matrix on two parties ${\cal H}_{A_\ta}^{\otimes n^{\tq-3}}$ and ${\cal H}_{A_\ta}^{\otimes n^{\tq-3}}$. This two-party classical density matrix is not properly normalized but we can normalize it to one because the inequalities that we are interested in checking do not depend on the norm of the state.
In this way, the proof of inequality $\rcg$ for $\tq-1$-party classical density matrix now reduces to that for $2$-party classical density matrix. Before moving to the general case, let us consider the case of two correlated coins for concreteness. As mentioned earlier, this is equivalent to a diagonal density matrix over two parties where each party is $2$-dimensional.

\subsection{Heads or tails}
In this subsection, we will consider a classical density matrix i.e. a completely diagonal density matrix over two parties, $\1$ and $\2$. As emphasized earlier, this is simply a two-variable probability distribution. Let us take the density matrix to be
\begin{align}
    \rho_{\alpha_\1 \alpha_\2}^{\beta_1 \beta_2} = p_{\alpha_\1\alpha\2} \,  \delta_{\alpha_\1}^{\beta_\1} \, \delta_{\alpha_\2}^{\beta_\2}.
\end{align}
We can think of $p_{\alpha_\1\alpha_\2}$ as the probability of outcomes $(\alpha_\1,\alpha_\2)$ in an experiment. Let us specialize to the case with the two parties being coins i.e. they both have only two states $0$ and $1$ each. So $\alpha_1$ as well as $\alpha_2$ take only two values $0$ and $1$. The measure $({\cal E}_n^{(\2)})^n$ is simply $n$-th power of the Renyi entropy for the combined event i.e.
\begin{align}\label{coin2}
    ({\cal E}_n^{(\2)})^n = (p^n_{00}+p^n_{01}+p^n_{10}+p^n_{11})^n.
\end{align}
The calculation of ${\cal E}_n^{(\3)}$ is a little more involved. Consider the two-dimensional lattice of $\rho$'s as shown in figure \ref{q3}. Because the density matrix is diagonal, we will fix $m$ rows (corresponding to the first coin) and $m'$ columns (corresponding to the second coin) to have the state $0$. The states in the remaining rows and columns are taken to be $1$. The contribution of this configuration to ${\cal E}_n^{(\3)}$ is $p_{00}^{mm'}p_{01}^{m(n-m')}p_{10}^{(n-m)m'}p_{11}^{(n-m)(n-m')}$.
There are $\,^nC_m \,^nC_{m'}$ such configurations. Summing over all of them, we get
\begin{align}\label{coin3}
    {\cal E}_n^{(\3)}&=\sum_{m=0}^n \sum_{m'=0}^n \,^nC_m \,^nC_{m'}(p_{00}^{mm'}p_{01}^{m(n-m')}p_{10}^{(n-m)m'}p_{11}^{(n-m)(n-m')})\notag\\
    &= \sum_{m=0}^n \,^nC_m (p^m_{00} p^{n-m}_{10}+p_{01}^m p_{11}^{(n-m)})^n.
\end{align}
In the second line, we have summed over $m'$ to obtain a more compact expression. On the other hand, the advantage of the form that is given in the first line is that it is manifestly symmetric in both parties. We would like to show that the quantity in equation \eqref{coin3} is smaller than the one in equation \eqref{coin2} for arbitrary positive numbers $p_{\alpha_1 \alpha_2}=1$ but because both quantities have the same homogeneity $n^2$ in $p_{\alpha_1\alpha_2}$'s this normalization does not matter.

The equation \eqref{coin3} is massaged as follows.
\begin{align}
    {\cal E}_n^{(3)} &= \sum_{m,k =0}^{n} {}^nC_k  {}^nC_{m} p_{10}^{nk} a^{mk} p_{11}^{n(n-k)} b^{m(n-k)} = \sum_{m,k =0}^{n} {}^nC_k p_{10}^{nk} p_{11}^{n(n-k)} b^{mn} \times {}^nC_{m} c^{mk}
\end{align}
where we have defined, $a = \frac{p_{00}}{p_{10}},  b = \frac{p_{01}}{p_{11}} \ \& \ c = \frac{a}{b}$. Similarly, equation \eqref{coin2} is massaged as, 
\begin{align}
    ({\cal E}_n^{(2)})^n = \left( p_{10}^n(1 + a^n) + p_{11}^n(1 + b^n)  \right)^n = \sum_{k=0}^n {}^nC_k p_{10}^{nk} p_{11}^{n(n-k)} \sum_{r = 0}^k \sum_{s = 0}^{n-k} {}^kC_r {}^{n-k}C_{s} a^{nr} b^{ns}
\end{align}
Changing the variables from $s$ to $m$ as $s= m-r$,
\begin{align}
    ({\cal E}_n^{(2)})^n = \sum_{k,m = 0}^n {}^nC_k p_{10}^{nk} p_{11}^{n(n-k)} b^{mn} \times \sum_{r = 0}^k {}^kC_r {}^{n-k}C_{m-r} c^{n r}.
\end{align}
The range of summation of $m$ is from $r$ to $r+n-k$ but we have extended it from $0$ to $n$ because summand contains $\,^{n-k}C_{m-r}$ which vanishes in the added part of the summation range. 
Hence, proving the inequality $\rcg$ reduces to proving
\begin{equation}
    \sum_{k,m = 0}^n {}^nC_k p_{10}^{nk} p_{11}^{n(n-k)} b^{mn} \Big( \sum_{r = 0}^k {}^kC_r {}^{n-k}C_{m-r} c^{n r} -  {}^nC_{m} c^{mk}\Big) \geq 0.
\end{equation}
As it turns out, the inequality holds for each term in the $k,m$ sum.  
We will now prove the inequality, 
\begin{equation}\label{reduced-ineq}
    \sum_{r = 0}^k {}^kC_r {}^{n-k}C_{m-r} c^{n r}  \geq  {}^nC_{m} c^{mk}.
\end{equation}
for all values of $m, k$ and $n$.
As a sanity check, let us notice that the above inequality is saturated for $c=1$. This can be seen by considering the equality $(1+x)^{k} (1+x)^{n-k}= (1+x)^n$ and matching coefficient of $x^m$ on both sides. 
This is as it should be because $c=1$ corresponds to a factorized density matrix.

We will prove the inequality \eqref{reduced-ineq} using the fact that arithmetic mean $\geq$ geometric mean ({\tt{am-gm}}). Consider the left-hand side as the sum of monomials of gamma with coefficient $1$. The {\tt{am-gm}} inequality for this collection of monomials is
\begin{align}
    \frac{1}{\,^n C_m}\sum_{r = 0}^k {}^kC_r {}^{n-k}C_{m-r} c^{n r} \geq \Big(c^{n \sum_{r= 0}^k r\,\, {}^kC_r {}^{n-k}C_{m-r} } \Big)^{\frac{1}{\,^n C_m}}.
\end{align} 
We now need to show that the right-hand side of the above inequality is $c^{mk}$ i.e. we need to show
\begin{align}\label{check-general}
    n \sum_{r= 0}^k  r\,\, {}^kC_r {}^{n-k}C_{m-r} =  \,^n C_m \, m \, k.
\end{align}
We have shown that the inequality \eqref{reduced-ineq} is saturated at $c=1$ because it corresponds to a factorized density matrix. In section \ref{sec-factorized}, we have also shown that inequality $\rcg$ continues to be saturated at first order in perturbation about the factorized density matrix. Then it must be that the inequality \eqref{reduced-ineq} should be saturated if we differentiate both sides at $c=1$. This gives the equation \eqref{check-general} needed to prove the inequality \eqref{reduced-ineq} for general states.

Alternatively, we can prove the equation \eqref{check-general} directly by following combinatorial manipulations:
\begin{align} \label{combinatorial}
    n \sum_{r = 0}^k\,  r\, \, {}^k C_r {}^{n-k} C_{m-r} &= n\, k\, \sum_{r = 1}^k  {}^{k-1} C_{r-1} {}^{(n-1) - (k-1)} C_{(m-1) - (r-1)}\notag \\ &= n\, k\, \sum_{s = 0}^{k-1}  {}^{k-1} C_{s} {}^{(n-1) - (k-1)} C_{(m-1) - s},
\end{align}
where $s = r - 1$. Now, note that as explained below the equation \eqref{reduced-ineq}, we also have the identity,
\begin{equation}
    \sum_{r = 0}^k {}^k C_r {}^{n - k} C_{m - r} = {}^n C_m.
\end{equation}
This finally simplifies the equation \eqref{combinatorial} to give,
\begin{align}
    n \sum_{r = 0}^k\, r\, \ {}^k C_r {}^{n-k} C_{m-r} &= n\, k\, \, {}^{n-1} C_{m-1} = {}^n C_m \, m\, k.
\end{align}

\subsection{Generalization to an arbitrary number of states}
We have reduced the proof of inequality $\rcg$ for the classical density matrix on any number of parties to that for only two parties but with arbitrary dimensions. In the previous section, we proved $\rcg$ for the case of two parties, each with dimension $2$. In this section, we will generalize this proof to two parties with arbitrary dimensions $d_1$ and $d_2$. 

Let us consider a sector of configurations on the $n\times n$ lattice with states $s_1,\ldots, s_n$ and $t_1,\ldots, t_n$ on the rows and columns respectively. 
Note that these are the \emph{values} the indices $\alpha_1$ and $\alpha_2$ take respectively. They are not indices themselves.
We will prove the inequality $\rcg$ by proving it in a given $(s_i,t_i)$ sector.
To begin with let us assume that $d_1, d_2 \geq n$ and take $s_i$'s as well as $t_i$'s to be distinct. Contribution of this sector of configurations to ${\cal E}_n^{(3)}$ is,
\begin{align}
    {\cal E}_n^{(3)}|_{\rm sector}= n! n! \prod_{i,j} p_{s_i,t_j}.
\end{align}
Here, the two factors of $n!$ up front come from the number of ways of arranging the distinct $s_i$ or $t_i$ on the $n \times n$ lattice.

Now, we consider $({\cal E}_n^{(2)})^n$, which has the same number of vertices as ${\cal E}_n^{(3)}$. To compute $({\cal E}_n^{(2)})^n$, we need to orient party $2$ edges along party $1$ edges. Let the strand of party $2$ that is along the strand $s_i$ of party $1$ be $t_{\sigma\cdot i}$ for some permutation $\sigma$. We need to compute the contribution for each $\sigma$ and then sum over $\sigma$. 
\begin{align}
    ({\cal E}_n^{(2)})^n|_{\rm sector} = \sum_\sigma n! \prod_{i} p^n_{s_i,t_{\sigma\cdot i}}.
\end{align}
The inequality is now proved using {\tt am-gm}. Consider $({\cal E}_n^{(2)})^n$ as sum of $n!$ terms, each corresponding to a permutation. 
\begin{align}
    ({\cal E}_n^{(2)})^n|_{\rm sector} \geq n!\prod_{\sigma} (n!\prod_{i} p^n_{s_i,t_{\sigma\cdot i}})^{1/n!}=n!n! \prod_{i, j} \Big((p^n_{s_i,t_j})^{(n-1)!}\Big)^{1/n!}={\cal E}_n^{(3)}|_{\rm sector}.
\end{align}
In the first equality, we have realized that the number of permutations that yield a given term with $\sigma\cdot i=j$ is $(n-1)!$.

Let us now allow for repetitions among the index values $s_i$'s and $t_i$'s.
It is useful to have a simple example in mind while doing the counting. Let's say, as values of $s_i$ we have $m_s$ heads and $n-m_s$ tails and as values of $t_i$ we have $m_t$ heads and $n-m_t$ tails.
Let $n_s$ and $n_t$ be the number of elements in the orbit of $s_i$'s and $t_i$'s under permutations. For the example at hand,
\begin{align}
    n_s= \,^nC_{m_s}, \qquad n_t=\,^n C_{m_t}.
\end{align}
With this definition, we have
\begin{align}
    {\cal E}_n^{(3)}|_{\rm sector}&=n_s n_t \prod_{i,j} p_{s_i,t_j},\qquad ({\cal E}_n^{(2)})^n|_{\rm sector} = \sum_{\sigma \in {\rm nontriv}} n_s \prod_{i} p^n_{s_i,t_{\sigma\cdot i}}.
\end{align}
Here $\sigma \in {\rm nontriv}$ stands for non-trivial permutations of $t_i$.
These are permutations that result in different assignments of heads and tails to the indices $t_i$. These are precisely $n_t$ in number.
Using the {\tt am-gm} inequality as before,
\begin{align}
    ({\cal E}_n^{(2)})^n|_{\rm sector} \geq n_t\prod_{\sigma\in {\rm nontriv}} (n_s \prod_{i} p^n_{s_i,t_{\sigma\cdot i}})^{1/n_t}.
\end{align}
We have obtained this inequality starting from the expression of $({\cal E}_n^{(2)})^n|_{\rm sector}= n_t\cdot {\mathtt {am}}$. As there are $n_t$ terms in the product over $\sigma$ and $n_s$ is independent of $\sigma$, it comes out of the product and the $n_t$-th root. We get
\begin{align}
    ({\cal E}_n^{(2)})^n|_{\rm sector} \geq n_t n_s \prod_{i} \prod_{\sigma\in {\rm nontriv}} \Big(p^n_{s_i,t_{\sigma\cdot i }}\Big)^{1/n_t}.
\end{align}
In order to prove the desired inequality, we will show
\begin{align}
    \prod_{j} p_{s_i,t_j} = \prod_{\sigma\in {\rm nontriv}} \Big(p^n_{s_i,t_{\sigma\cdot i }}\Big)^{1/n_t}.
\end{align}
Let us evaluate the left-hand side for our example. In that case, $m_t$ number of $t_j$'s take the values of heads and $n-m_t$ number of $t_j$'s take the values tails. So 
\begin{align}\label{lhs}
    \prod_{j} p_{s_i,t_j} = (p_{s_i,{\rm heads}})^{m_t} (p_{s_i,{\rm tails}})^{n-m_t}.
\end{align}
On the right-hand side, different factors of $p_{s_i, t_j}$ are produced as we vary $\sigma$ because there will be $\sigma$'s such that $j=\sigma\cdot i$. The number of such $\sigma$'s is the size of the stabilizer of a given $j$. In our example, we fix one of the $t_j$ to be, say heads. Then the number of permutations among $n_t$ number of permutations that preserve this $t_j$ are 
\begin{align}\label{heads-fixed}
    \,^{n-1}C_{m_t-1} = \,^nC_{m_t}\,  \frac{m_t}{n}= n_t \frac{m_t}{n}.
\end{align}
This set of permutations yields only those values of $j$ for which $t_j$ is fixed to be heads. Using equation \eqref{heads-fixed} and the right hand side of \eqref{lhs}, their contribution is $(p_{s_i,{\rm heads}})^{m_t}$. A similar calculation for the tails gives those values of $j$ for which $t_j$ is fixed to be tails. Their contribution is $(p_{s_i,{\rm tails}})^{n- m_t}$. Multiplying these contributions, we precisely get the contribution of the equation \eqref{lhs}. 

We can generalize this argument to any number of outcomes. The reason behind the presence of binomial coefficients in the expressions above is that we are restricting ourselves to the case where $s_i$ and $t_i$ ($i = 1, 2, ..., n$) take either of the two values (heads or tails). When we generalize to the case where $s_i$ and $t_i$ can take more than two values, the binomial coefficients will get replaced by appropriate multinomial coefficients. For a set of $l$ distinct outcomes with $k$-th outcome appearing $m_k$ number of times such that $\sum_{k=1}^{l} m_k= n$, we get
\begin{align}
    n_t= \frac{n!}{m_1!\ldots m_l!}.
\end{align}
The counting argument remains unaffected.

\subsection{What is Renyi multi-entropy?}
We end the section on classical states by describing a somewhat amusing interpretation of ${\cal E}_n^{(3)}$ for probability distributions for two correlated coins.

Let us first consider the meaning of ordinary ${\cal E}_n^{(2)}$ for the probability distribution for a single coin. Let the probability for heads and tails be $p_0$ and $p_1$ respectively. The quantity 
\begin{align}
    {\cal E}_n^{(2)}= p_0^n+p_1^n
\end{align}
is the probability of observing the same outcome in $n$ coin tosses. 

To understand the meaning of ${\cal E}_n^{(3)}$ let us consider two sets of $n$ people each $\{r_1,\ldots, r_n \}$ and $\{b_1,\ldots, b_n\}$. Let the group of people $\{r_i\}$ have a coin $R$ and the group of people $\{b_i\}$ have the coin $B$. The measure $({\cal E}_n^{(2)})^n$ can be thought of as follows. Each person $r_i$ \emph{picks a partner}, say $b_i$ from the other group. They pair up and toss their coins $n$ times. This happens for each $i$. All in all, there is $n^2$ number of experiments.
The probability of each person observing the same result for their own coin is precisely  $({\cal E}_n^{(2)})^n$.

Now consider a slightly different experiment. Each person $r_i$ now pairs up with \emph{every person} $b_j$ in the other group, one after the other,  to perform the tossing of the $R, B$ coins. Again there is $n^2$ number of experiments.  The probability of each person observing the same result for their own coin is now  $({\cal E}_n^{(3)})$. This viewpoint can be generalized to two-party classical density matrices over higher dimensional spaces i.e. to the case of ``multi-sided dice'' straightforwardly. However, it seems difficult to generalize this interpretation for ${\cal E}_n^{(\tq)}$ with $\tq\geq 4$.

\section{Generalized W state}\label{genw}

\begin{figure}[t]
    \begin{center}
        \includegraphics[scale=0.8]{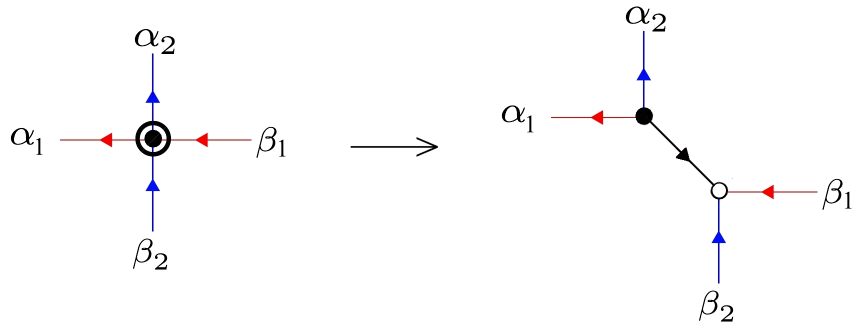}
    \end{center}
    \caption{Resolution of the 4-valent density matrix.}\label{resol}
\end{figure}

\begin{figure}[t]
    \begin{center}
        \includegraphics[scale=0.2]{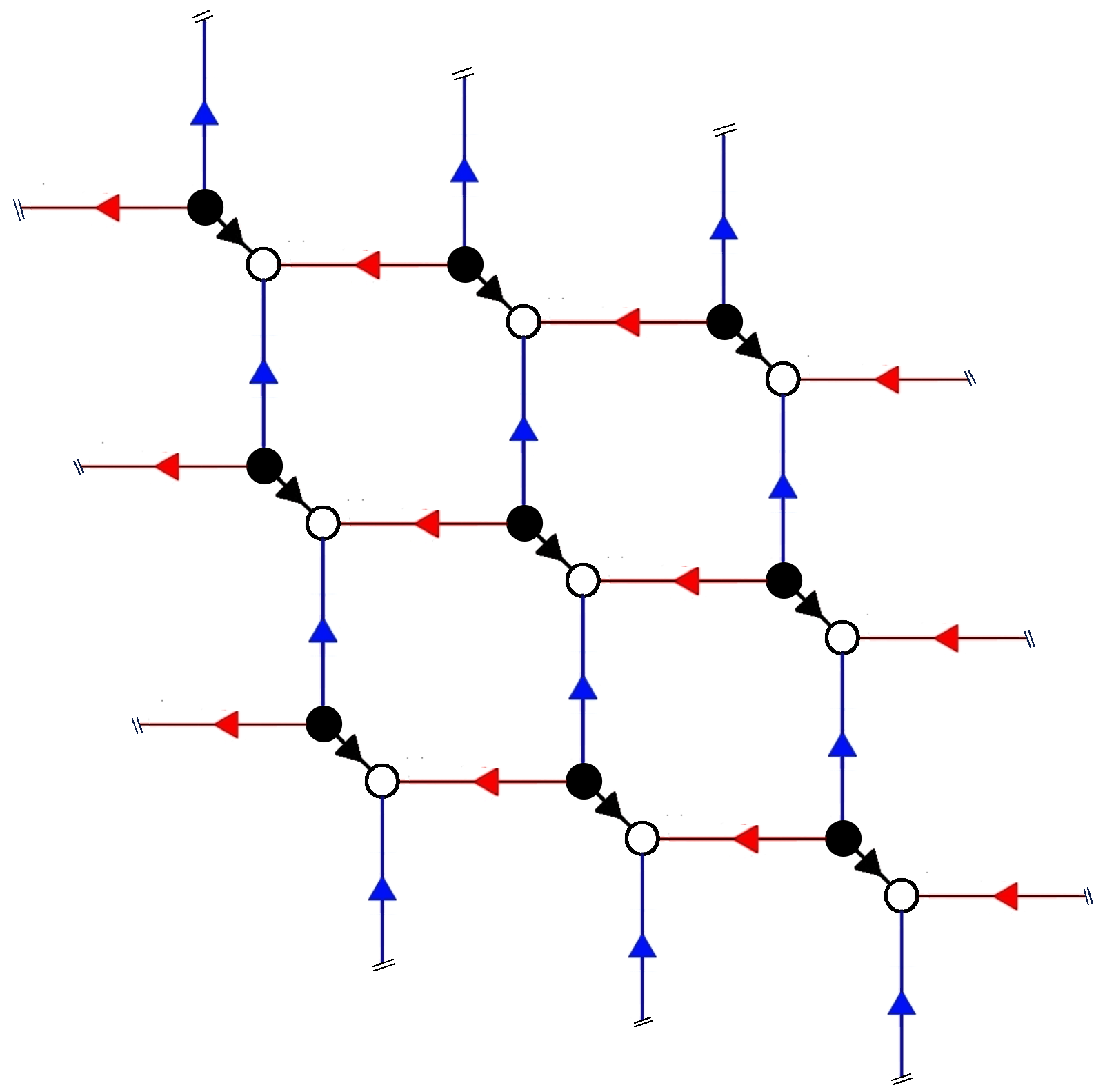}
    \end{center}
    \caption{Hexagonal lattice obtained after resolution for $n = 3$.}\label{hexa}
\end{figure}

In this section, we will compute Renyi multi-entropy for a special type of three-qubit state that is a generalization of the so-called W-state and check $\scg$. We call the state with the following form to be the generalized W-state:
\begin{align}
    |W\rangle = c_1|100\rangle + c_2|010\rangle +c_3|001\rangle.
\end{align}
When all the $c$'s are taken to be the same, the state is known as the W-state. We will think of Renyi multi-entropy as the partition function of the statistical mechanical system as outlined in section \ref{lattice}. However, instead of tracing out one party and thinking of the resulting density matrix as a 4-valent vertex in a square lattice, it helps to think in terms of the 3-valent vertex of the state itself. When we ``resolve'' the 4-valent density matrix into the state and its complex conjugate, we get a pair of 3-valent vertices, one black and one white. This resolution is shown in figure \ref{resol}. When the vertices of the square lattice are resolved in this way, we get a hexagonal lattice with periodic boundary conditions as shown in figure \ref{hexa}. A nice property of this lattice is that, unlike the square lattice, it is manifestly symmetric in all the parties. 

To compute the Renyi multi-entropy as the partition function on this lattice, the Boltzmann weight at each vertex is taken to be the state itself. We now have to sum over all configurations of the edges. An edge can either take the value $|0\rangle$ and $|1\rangle$. If we interpret $|1\rangle$ as occupied and $|0\rangle$ as vacant, the problem precisely maps to the fully packed dimer model on the hexagonal lattice. The edges of the lattice are distinguished into three types depending on their spatial orientation. The Boltzmann weight at each vertex dictates we weigh the empty edge by $1$ and the occupied edge by $z_\ta\equiv |c_\ta|^2$ depending on the orientation.  The fully packed dimer model on any planar lattice was solved by Kasteleyn around 60 years ago \cite{KASTELEYN19611209, 10.1063/1.1703953}. He also figured out how to deal with non-planarity arising from imposing periodic boundary conditions. We review his formalism briefly in appendix \ref{kast}. Here we directly start from the result.

The quantity ${\cal E}_n^{(3)}$ for W-state is given as,
\begin{align}\label{3partiteW}
    {\cal E}_n^{(3)}(|W\rangle) &=  \frac{1}{2} \left( \sZ_{+-} +\sZ_{-+}  + (-1)^{n} \left(\sZ_{--}-\sZ_{++} \right) \right)  \\
\textrm{where, } \quad \sZ_{s_1 s_2} &= 
\prod_{i=0}^{n-1} ((z_1+  \omega^{i+\frac{1-s_1}{4}} z_2)^n-(-\omega^{\frac{1-s_2}{4}})^n  z_3^n), \quad \quad \omega = e^{\frac{2 \pi i}{n}}, s_{1,2}= \pm. \notag
\end{align} 
Several comments are in order regarding the equation \eqref{3partiteW}. First, although it is not manifest, ${\cal E}_n^{(3)}$ is symmetric under permutations of  $z_1,z_2$ and $z_3$. Let us write $Z_{++}$ as sum of monomials,
\begin{align}\label{zpp}
    Z_{++}=\sum_{r_1,r_2,r_3} c_{r_1 r_2 r_3} z_1^{r_1} z_2^{r_2} z_{3}^{r_3}.
\end{align}
The coefficient $c_{r_1 r_2 r_3}$ is real but does not have a definite sign. In fact, the particular combination of $Z_{s_1 s_2}$ that appears in ${\cal E}_n^{(3)}$ is constructed precisely so that the expansion of ${\cal E}_n^{(3)}$ in the monomials is the same as \eqref{zpp} but with $|c_{r_1 r_2 r_3}|$ replacing $c_{r_1 r_2 r_3}$. See section \ref{kast} for an explanation. 

To prove the inequality $\scg$ we also want to compute ${\cal E}_n^{(2)}$ after identifying two parties. Without loss of generality, we take those two parties to be qubit $1$ and $2$. We get,
\begin{equation}
    {\cal E}_n^{(2)}(|W\rangle) |_{[12]} =  z_3^{n} + (z_1 + z_2)^{n}.
\end{equation}
We would now like to show ${\cal E}_n^{(3)}(|W\rangle) \leq ({\cal E}_n^{(2)}(|W\rangle))^n$ for all values of $z_\ta \geq 0$. For a normalized state, we have $z_1+z_2+z_3=1$ but we can ignore the normalization in proving the inequality because both sides have the same overall homogeneity in $z_\ta$.  Let us compare a single term in the product in $Z_{++}$ with ${\cal E}_n^{(2)}$. It is clear that 
\begin{align}\label{w-ineq}
    |((z_1+\omega^i z_2)^n+(-)^n z_3^n)|_{z_1^{r_1} z_2^{r_2} z_{3}^{r_3}}|  \leq (z_3^{n} + (z_1 + z_2)^{n})|_{z_1^{r_1} z_2^{r_2} z_{3}^{r_3}}.
\end{align}
Here the subscript $|_{z_1^{r_1} z_2^{r_2} z_{3}^{r_3}}$ denotes the coefficient of the monomial $z_1^{r_1} z_2^{r_2} z_{3}^{r_3}$. The above inequality \eqref{w-ineq} is true because of the potential destructive interference on the left-hand side. On the right-hand side, all the terms appear with the positive sign so there is no interference. Now taking the product over $i$ on both sides (the right-hand side is independent of $i$) gives us our desired inequality.

\section{Holographic states}\label{holo}

A brane-web prescription for computing Renyi multi-entropy for holographic CFTs was given in \cite{Gadde:2022cqi} which was in turn inspired by Lewkowycz and Maldacena's derivation \cite{Lewkowycz:2013nqa} of the Ryu-Takayanagi formula \cite{Ryu:2006bv}. An important assumption that went into deriving this prescription is that the dominant bulk solution dual to the measure preserves replica symmetry. It is not clear whether this assumption is true in general.  But for holographic  CFTs in two dimensions and for $n=2$ it is true. This was shown in the case of three intervals in \cite{Penington:2022dhr} and in the case of four intervals in \cite{Gadde:2023zzj}. We expect this to be true for any number of intervals \cite{ebk01:3460000000134968, GaddeWIP}. In this section, we will use the brane-web prescription to argue that the inequality $\scg$ with $n=2$ holds for holographic CFTs in $2d$. Before we get into the proof, let us review the results of \cite{Gadde:2022cqi}.

Although the results of \cite{Gadde:2022cqi} and the proof of $\scg$ presented here are valid for a wide range of states of a $2d$ holographic  CFT defined at the moment of time symmetry,  such as the thermo-field double state or multi-boundary wormhole states, in the following presentation we will only consider the CFT vacuum state on a circle for simplicity.  Let ${\cal R}_\ta$, $\ta=\1,\ldots,\tq$ be $\tq$ not-necessarily-connected regions on the boundary at a time-symmetric Cauchy slice ${\cal R}$. The measure ${\cal E}_n^{(\tq)}$ is computed as a partition function on a sphere that is ramified at the endpoints of the distinct boundary regions. This ramification is obtained by considering multiple replicas of the $2d$ theory on the sphere, cutting all of them along ${\cal R}$ and gluing the regions ${\cal R}_\ta$ according to the permutation elements $\sigma_\ta$ given in section \ref{mrenyi-def}. In \cite{Gadde:2022cqi}, the free energy on this manifold was computed using holography as the classical action of the dual gravitational solution ${\cal B}_n$ by assuming that this solution preserves replica symmetry. As remarked earlier, it is not clear whether the replica symmetry assumption is true in general but for $n=2$ it is true \cite{ebk01:3460000000134968, GaddeWIP}. Here we will not give the details of the holographic computation but simply state the result and highlight some of its relevant features. For details of the derivation, see \cite{Gadde:2022cqi,Gadde:2023zzj}.  

The gravitational action of the dual solution ${\cal B}_n$ is $n^{\tq-1}$ times the gravitational action of the conical geometry ${\widetilde {\cal B}}_n$. The conical geometry is locally AdS with a spherical boundary but with a locus of co-dimension $2$ conical singularities with cone angle $2\pi/n$. We will refer to $n$ as the cone parameter.  The singular locus is constrained to end on the ramification points on the boundary sphere.  It forms a tri-valent tree graph ${\cal W}$ which, because of the ${\mathbb Z}_2$ time reflection symmetry lies entirely in the bulk time symmetric Cauchy slice ${\cal C}$ such that $\partial {\cal C}={\cal R}$. 
The nice thing about the geometry ${\widetilde {\cal B}}_n$ is that it can be analytically continued as a cone manifold for any real value of $n$ between $1$ and $2$ \cite{ebk01:3460000000134968, GaddeWIP}. For future convenience, let us note that the trivalent tree decomposes the bulk Cauchy slice ${\cal C}$ into chambers. Each of the chambers is adjacent to a particular region on the boundary. We will label the union of all the chambers that are adjacent to the boundary region ${\cal R}_\ta$ (which could be a union of multiple connected intervals) as ${\cal C}_\ta$. After this labeling, no chamber within ${\cal C}$ remains unlabeled. 
As remarked earlier, the measure ${\cal E}_2^{(\tq)}$ can be calculated in the limit of large central charge as the action,
\begin{align}
    {\cal E}_2^{(\tq)}= \exp\Big(-n^{\tq-1} {\cal S}_{\rm grav}({\widetilde {\cal B}}^{(\tq)}_2)\Big).
\end{align} 
Here we added the superscript $(\tq)$ on ${\widetilde {\cal B}}_n$ to emphasize that this cone geometry is dual to a given $\tq$-partite decomposition of the boundary and ${\cal S}_{\rm grav}$ is the gravitational action. Now we identify two parties on the boundary, say $\ta_0$ and $\tb_0$ and consider the corresponding $\tq-1$ partite measure ${\cal E}_2^{(\tq-1)}$. It is computed by the action of the new cone geometry obtained after modifying the topology of the singular locus appropriately. The modification is as follows. If there is a piece ${\cal W}_{\ta_0,\tb_0}$ of the singular locus that separates any of the components of ${\cal C}_{\ta_0}$ and ${\cal C}_{\tb_0}$ in ${\widetilde {\cal B}}_n$, then it is to be removed in ${\widetilde {\cal B}}^{(\tq-1)}_n$. If there is no such piece then the new geometry is identical to the old one. 

\begin{figure}[t]
    \begin{center}
        \includegraphics[scale=0.7]{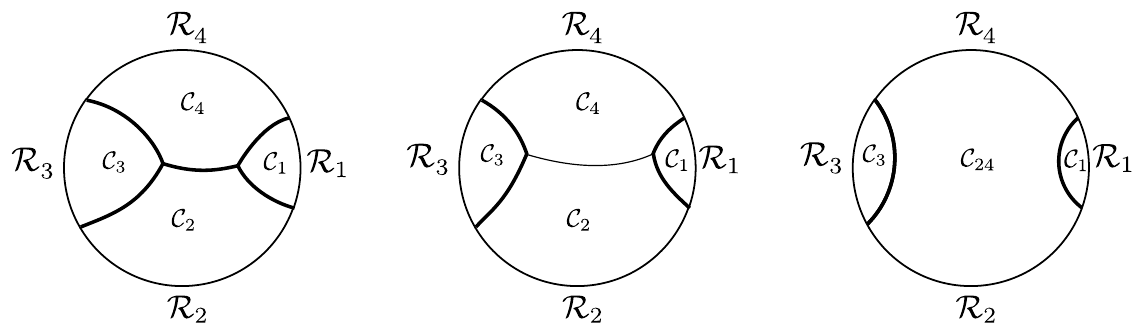}
    \end{center}
    \caption{The evolution of the web $\mathcal{W}$ corresponding to $\mathtt q = 4$ as the value of $n_{2,4}$ changes from $2$ to $1$. The other cone parameters $n_{\ta,\tb}$ are fixed to $2$. The thickness of the segment indicates the value of the cone parameter.}\label{holographic}
\end{figure}

In order to prove the inequality $\scg$, we then need to show ${\cal S}_{\rm grav}({\widetilde {\cal B}}^{(\tq)}_2) \geq  {\cal S}_{\rm grav}({\widetilde {\cal B}}^{(\tq-1)}_2)$. As remarked earlier, the parameter $n$ at each of the singular segments can take any value between $1$ and $2$. Let us denote the cone parameter $n$ at the segment ${\cal W}_{\ta_0,\tb_0}$ by $n_{\ta_0,\tb_0}$ and denote the corresponding solution, with cone parameters at all other segments fixed to $2$, as ${\widetilde {\cal B}}^{(\tq)}_2(n_{\ta_0,\tb_0})$. As we change $n_{\ta_0,\tb_0}$ from $2$ to $1$, the geometry ${\widetilde {\cal B}}^{(\tq)}_2$ smoothly goes over to ${\widetilde {\cal B}}^{(\tq-1)}_2$ because we have ${\widetilde {\cal B}}^{(\tq)}_2(1)={\widetilde {\cal B}}^{(\tq-1)}_2$. The desired inequality can now be obtained by proving 
\begin{align}\label{der-pos}
    \partial_{n_{\ta_0,\tb_0}}{\cal S}_{\rm grav}({\widetilde {\cal B}}^{(\tq)}_2(n_{\ta_0,\tb_0})) \geq 0.
\end{align}
This strategy is illustrated in figure \ref{holographic}. In the figure, we have considered the case of $\tq=4$ where the parties correspond to the regions ${\cal R}_\ta, \ta=1,\ldots, 4$. The $\tq-1$ partite state is obtained by identifying the parties $\ta_0=\2$ and $\tb_0=\4$. The cone parameter at the segment of the web that separates the chambers ${\cal C}_\2$ and ${\cal C}_\4$ is taken continuously from $2$ to $1$. 

The derivative of the gravitational action with respect to $n$ was studied in \cite{Dong:2016fnf} for the bi-partite case (see also \cite{Dong:2017xht}) and it was shown that
\begin{align}\label{mod-dong}
    \partial_n S({\widetilde {\cal B}}_n^{(\2)})= \frac{1}{n^2} \frac{1}{4G_N} A({\cal W}).
\end{align}
Here $A({\cal W})$ is the area of the singular locus ${\cal W}$ with conical angle $2\pi/n$ anchored at the ramification points on the boundary. For the bi-partite case, the locus does not have junctions but is formed by disjoint segments joining boundary points. We need a generalization of this formula to the case where the singular locus has junctions and the cone parameters in each segment can be varied independently. Such a general formula exists and is known in the mathematics literature as the Schlafli formula for cone manifolds \cite{ebk01:3460000000134968}. It takes the form,
\begin{align}\label{schlafli}
    \partial_t S({\widetilde {\cal B}}_n^{(\2)})= \sum_{\ta, \tb }\frac{1}{n_{\ta,\tb }^2} \frac{1}{4G_N} A({\cal W}_{\ta,\tb}) \frac{dn_{\ta,\tb}}{dt}.
\end{align}
Here ${\cal W}_{\ta,\tb}$ denotes the segment of ${\cal W}$ that separates the chambers ${\cal C}_\ta$ and ${\cal C}_{\tb}$. If these chambers are not adjacent then  ${\cal W}_{\ta,\tb}$ is empty. Thus the sum $\sum_{\ta,\tb}$ is simply the sum over all the segments of ${\cal W}$. The cone parameters $n_{\ta,\tb}$ at ${\cal W}_{\ta,\tb}$ are taken to be independent  functions of some parameter $t$. By taking $n_{\ta_0,\tb_0}=t$ and all other $n_{\ta,\tb}$'s to be $2$ we have the desired result \eqref{der-pos}.

The Schlafli formula \eqref{schlafli} is not difficult to understand from a physics point of view. 
Let us first review the proof of equation \eqref{mod-dong}, as given in \cite{Dong:2016fnf}. A co-dimension $2$ cosmic brane of tension $T= \frac{n-1}{n}\frac{1}{4G_N}$ produces around itself a conical singularity with conical angle $2\pi /n$. This is exactly the conical angle at the singular locus ${\cal W}$ of the orbifold geometry ${\widetilde {\cal B}}$. So we use coupling of gravity to such a cosmic brane as a trick to engineer required singularity at ${\cal W}$.  More explicitly, we consider a coupled system with the action 
\begin{align}
    {\cal S}_{\rm tot}={\cal S}_{\rm grav}+\sum_{\ta,\tb}{\cal S}_{\rm br}(T_{\ta,\tb}), \qquad\qquad {\cal S}_{\rm br}(T_{\ta,\tb}) \equiv T_{\ta,\tb} \int dy \sqrt{h},\,\, T_{\ta,\tb}=\frac{n_{\ta,\tb}-1}{n_{\ta,\tb}}\frac{1}{4G_N}.
\end{align}
Here the brane action ${\cal S}_{\rm br}$ is supported over a web that has the same topology as the singular locus. The segment is labeled by $(\ta,\tb)$ if it separates the chamber ${\cal C}_{\ta}$ and ${\cal C}_{\tb}$.  The brane supported on this segment is taken to have the tension $T_{\ta,\tb}$.

After solving the equations of motion for both the brane and the gravity, we precisely get ${\widetilde {\cal B}}_n$ with the brane supported over the singular locus ${\cal W}$. To compute $S_{\rm grav}$ on the resulting geometry we drill out the neighborhood of the conical singularity. 
By virtue of being a solution to the equation of motion, a first-order variation of the bulk gravity action with respect to metric variation is zero. The change is supported only at the boundary of the drilled-out tube. This is precisely negative of the variation of the Gibbons-Hawking-York (GHY) boundary term.
Taking the metric variation to correspond to the change in the conical angle, the contribution of this boundary term was evaluated in \cite{Dong:2016fnf} and was shown to be given by the formula \eqref{mod-dong}. 

To prove \eqref{schlafli}, we consider the case where the singular locus forms a tri-valent graph with its edges having any cone parameter $n\in [1,2]$ and drill out a tubular neighborhood of this graph. 
Again, by virtue of being a solution to the gravitational equation of motion, the variation induced by $\partial_t$ is supported purely on the boundary of the tube. The contribution coming from the tubular neighborhood of a single edge is exactly as given in \cite{Dong:2016fnf}. Summing over all the edges we get the Schlafli formula. {\emph A priori} there can be a contribution supported at the junction i.e. at the vertices of the graph ${\cal W}$. It was argued in \cite{Gadde:2022cqi, Gadde:2023zzj} that the junction contribution vanishes.  

Let us point out a peculiar property of the holographic states with the help of our measure. In the proof of the coarse graining inequality, we showed how one can interpolate from the old solution corresponding to $S_2^{(\tq)}$ to the new solution corresponding to $S_2^{(\tq-1)}$ obtained by identifying the parties $\ta_0$ and $\tb_0$. If there is a segment of the singular locus separating the chambers ${\cal C}_{\ta_0}$ and ${\cal C}_{\tb_0}$ in the solution corresponding to the $\tq$-partite state, it is removed in the solution corresponding to the $\ta-1$ partite state. However, the regions ${\cal R}_{\ta_0}$ and ${\cal R}_{\tb_0}$ may be ``sufficiently far'' from each other that the corresponding chambers don't share a wall. In that case $S_2^{(\tq)}$ is the same as $S_2^{(\tq-1)}$ and the coarse graining inequality is saturated. This makes the measure $S_2^{(\tq)}$ particularly well-suited for holographic states.
 
\section{Numerical checks}\label{numerical}

In this section, we summarize the numerical checks of the inequality $\scg$. We have computed the quantities on both sides of $\scg$ numerically for randomly chosen states for various values of $\tq$ and $n$. These computations are done in {\tt {Julia}} using the tensor index manipulation package {\tt {ITensor}}. Both time and memory for these computations scale faster than exponentially with the number of parties $\tq$ and $n$. 
We observed that the inequality $\scg$ holds for all the states checked. 
Let us denote the set of randomly chosen $\tq$-partite states with Hilbert space dimensions $d_\ta, \ta=\1,\ldots, \tq$ as $[d_\1,\ldots, d_\tq]$. Then we checked the inequality numerically for
\begin{enumerate}
    \item $\tq=3$: 
    \begin{itemize}
        \item $10^6$ states in $[2,2,16]$ under identification of any pair of parties up till $n=14$.
        \item $10^6$ states in $[3,3,16]$ under identification of any pair of parties up till $n=8$.
    \end{itemize}

    \item $\tq=4$: 
    \begin{itemize}
        \item $10^5$ states in $[2,2,2,16]$ under identification of any pair of parties up till $n=8$.
        \item $10^5$ states in $[2,3,3,16]$ under identification of any pair of parties up till $n=4$.
    \end{itemize}
        
    \item $\tq=5$: 
    \begin{itemize}
        \item $10^5$ states in $[2,2,2,2,16]$ under identification of any pair of parties up till $n=4$.
    \end{itemize}
\end{enumerate}
Increasing the value of dimensions $d_\ta$ is also numerically intensive. However, we have checked the inequality for higher values of $d_\ta$'s but for fewer states.

\section{Summary and outlook}\label{discuss}

In this paper, we have studied the Renyi multi-entropy $S_n^{(\tq)}$ for a diverse class of states. These multi-partite measures were introduced in \cite{Gadde:2022cqi,Gadde:2023zzj} motivated by their potential application to holography. In this paper, however, we focused on their quantum information-theoretic properties. We conjectured that the  $S_n^{(\tq)}$ is monotonic under coarse graining and proposed that this property makes it a good measure of multi-partite entanglement. We provided evidence for this conjecture by explicitly verifying the monotonicity for various classes of states. An obvious line of investigation now is to either prove the monotonicity conjecture or to find a counter-example. We plan to pursue this line in the future. 

The definition \eqref{m-renyi-def} of $S_n^{(\tq)}$ requires that the quantity ${\cal E}_n^{(\tq)}$ be positive definite. In this paper, we did not prove or even discuss this and phrased the monotonicity conjecture in terms of ${\cal E}_n^{(\tq)}$ instead. Although we found that ${\cal E}_n^{(\tq)}$ is positive in the examples we studied, it would be nice to prove this in general and to see if there is any relation between the positivity of ${\cal E}_n^{(\tq)}$ and the coarse-graining monotonicity. 

In addition to being monotonic under coarse graining, a good measure of multi-partite entanglement must also be monotonic under {\tt {locc}}. We did not investigate this aspect of the measure at all in this paper. In this context, it is worth mentioning the following. As stated in section \ref{intro}, a measure that is monotonic under {\tt{locc}} must be constant on separable mixed states. The pure state measure discussed in this paper is a smooth function of the state. Such a function can not be constant on a certain continuous family of states namely separable states and nontrivial on others. That is why the definition of an {\tt{locc}} monotonic entanglement measure for mixed states always involves some sort of optimization. 
In \cite{Uhlmann1998}, Uhlmann developed a general way of extending pure state entanglement measures to mixed states via the so-called ``convex roof construction''. This construction was used before in \cite{Bennett:1996gf} to extend the von Neumann entropy for bi-partite pure states to the ``entanglement of formation'' for mixed states. In \cite{Vidal:1998re}, Vidal showed that if a measure on the bi-partite pure state is concave in the single party density matrix obtained by tracing out one of the parties then its convex roof extension to mixed states is an {\tt{locc}} monotone. It is conceivable that a version of Vidal's proof holds also for convex roof extensions of multi-partite entanglement measures. If so, the question of whether our measure is concave in the density matrix becomes important. We hope to investigate this question in the future.

On another note, our proof of coarse-graining monotonicity for holographic states used the Schlafli formula but not in its full generality. The Schlafli formula allows for the cone parameters at other segments to be fixed to any value $\in[1,2]$, not just to $2$, as we take the cone parameter of one of the segments to $1$. A trivalent graph of singularities with possibly different values of cone parameters at every segment is precisely what we get when we consider a larger family of entanglement measures called the ``special symmetric measures'', introduced in \cite{Gadde:2023zzj}. In section \ref{mrenyi-def}, we have labeled general local unitary invariants of a $\tq$-partite state by a $\tq$-tuple of permutation elements $\sigma_\ta$. The special symmetric measures correspond to choosing $\sigma_\ta$ to be the generators of ${\mathbb Z}_{m_\ta}$ with replicas taken to be in a $m_\1\times\ldots m_\tq$ hyper-cubical lattice with ${\mathbb Z}_{m_\ta}$ acting as a lattice translation in direction $\ta$. 
The Renyi multi-entropy studied here is a member of this family. When $m_\ta$ are all taken to be equal to $n$, it reduces to (the $n$-th power of) ${\cal E}_n^{(\tq)}$. See \cite{Gadde:2023zzj} for details. The monotonicity property following from the general form of the Schlafli formula suggests a tantalizing possibility that the entire family of special symmetric measures is monotonic under coarse-graining. It would be nice to check this conjecture, or better yet, to prove it. It would also be good to analyze its concavity property and convex roof extension. If these measures turn out to have monotonicity under coarse-graining as well as under {\tt{locc}} then it is possible that they form a \emph{complete} set of good multi-partite entanglement measures capable of distinguishing multi-partite states up to local unitary transformations. 

It is also instructive to study the saturation of the coarse-graining inequalities. As pointed out towards the end of section \ref{holo}, this is particularly important for holographic states as they saturate a number of coarse graining inequalities. This saturation corresponds to the vanishing of a multi-partite version of mutual information. It would be nice to understand the precise sense in which this is true. The vanishing of ordinary mutual information plays an important role in diagnosing correctable errors via the so-called ``decoupling principle'' \cite{PhysRevA.54.2629}. This diagnosis of quantum error correction has played an important role in understanding bulk reconstruction \cite{Dong:2016eik, Almheiri:2014lwa, Harlow:2016vwg}. If we can understand how the saturation of the coarse-graining inequalities is related to quantum error correction for multiple parties, it would give a refinement of the local AdS/CFT correspondence as discussed in the last section of \cite{Gadde:2023zzj}. 

\section*{Acknowledgements}
We would like to thank Gautam Mandal, Arvind Nair, Onkar Parrikar, Pratik Rath, Pranab Sen, Piyush Shrivastava, Sandip Trivedi for interesting discussions. We are particularly indebted to Shiraz Minwalla for his insightful comments.
This work is supported by the Infosys Endowment for the study of the Quantum Structure of Spacetime and by the SERB Ramanujan fellowship.  We acknowledge the support of the Department of Atomic Energy, Government of India, under Project Identification No. RTI 4002. HK would like to thank KVPY DST fellowship for partially supporting his work.
Finally, we acknowledge our debt to the people of India for their steady support to the study of the basic sciences.

\appendix
\section{Kasteleyn's formalism}\label{kast}
Here we briefly review techniques used by Kasteleyn to compute the partition function of the fully packed dimer model on a graph. Let us first define some terminology. Let $L$ be a lattice with an even number $2m$ of vertices and $R$ be the set of its edges. Let us assume that $R$ is divided into several classes $C_1,\ldots, c_h$. As an example, we can consider a square lattice in two dimensions, then the set of vertical bonds and horizontal bonds form two different classes. In the case of regular hexagonal or honeycomb lattice, the bonds are naturally divided into three classes by their angle with respect to, say the positive horizontal axis. We can take the bonds to be aligned at angles $0$, $2\pi/3$ and $4\pi/3$ with respect to this axis. For the generalized $W$ state discussed in section \ref{genw}, this is the relevant lattice model. A dimer can be placed on any of the bonds but with the constraint that two dimers can not overlap at a vertex. A fully packed configuration of dimers is one where every single vertex is occupied by a dimer. Kasteleyn developed a theory of computing the partition function of the fully packed dimer model. Let the number of fully packed dimer configurations where $N_i$ bonds of the type $C_i$ occupied by dimers be $g_L(N_1,\ldots, N_h)$. Then the partition function of the fully packed dimer model on $L$ is
\begin{align}
    Z_{\rm dimer}(z_1,\ldots, z_h)= \sum_{\rm config} g_{L}(N_1,\ldots, N_h) z_1^{N_1}\ldots z_h^{N_h}
\end{align}
where the sum is taken over all fully packed dimer configurations. The measure ${\cal E}_n^{(3)}$ for the generalized W state considered in section \ref{genw}, is precisely the partition function $Z(z_1,z_2,z_3)$ on a hexagonal lattice with periodic boundary conditions where $z_a$ is the fugacity for each of the three types of bonds. Its relation to the parameters $c_a$ of the state is $z_a= |c_a|^2$.

Let us assign an orientation on the lattice $L$ by putting arrows on all the edges. Corresponding to this orientation we define an \emph{antisymmetric} adjacency matrix $M_{ij}$ such that $M_{ij}=z_a$ if the edge between vertices $i$ and $j$ is of class $C_a$ and is oriented \emph{from} $i$ \emph{to} $j$.  Because of anti-symmetry, $M_{ji}=-M_{ij}$ and $M_{ii}=0$. The Pfaffian of any  anti-symmetric matrix $M$ is defined as
\begin{align}
    {\rm Pf}(M)& = \sum\nolimits^{'}_{\sigma\in S_{2m}}{\rm sgn}_\sigma\,\, M_{\sigma\cdot 1, \sigma\cdot 2}\ldots M_{\sigma\cdot (2m-1), \sigma\cdot (2m)}\label{pf}\\
    &= \frac{1}{2^m m!}\sum\nolimits_{\sigma\in S_{2m}} {\rm sgn}_\sigma \,\, M_{\sigma\cdot 1, \sigma\cdot 2}\ldots M_{\sigma\cdot (2m-1), \sigma\cdot (2m)}.\notag
\end{align}
where the sum in the first equation is taken over only those permutations $\sigma$ that obey $\sigma\cdot ({2i-1}) < \sigma \cdot (2i)$ for any $i$ and $\sigma \cdot 1 <\sigma \cdot 3<\sigma \cdot 5 \ldots$. The sum in the second line is unconstrained. The two lines are equal because $M_{ij}$ is an antisymmetric matrix. The Pfaffian has the property that $|{\rm Pf}(M)|^2 = {\rm Det}(M)$. It is not difficult to see that when $M$ is taken to be the antisymmetric adjacency matrix of $L$, each term in the summation in equation \eqref{pf} is the contribution to the partition function of a fully packed dimer configuration. But due to the factor ${\rm sgn}_\sigma$ and the antisymmetry of $M$, the term comes with either with positive or with negative sign. In the partition function $Z$, we need to count each term with the same sign. The question is then the following: does there exist an orientation on $L$ such that each term in the ${\rm Pf}(M)$ comes with the same sign? Kasteleyn solved this problem in $1961$ \cite{KASTELEYN19611209, 10.1063/1.1703953}. He showed that such an orientation exists if the genus of the lattice graph $L$ is zero. By genus of the graph, we mean the minimum genus of the Riemann surface on which the graph can be drawn without crossing.  He also showed that if the genus is $g$ then the partition function can be written as a sum over $2^{2g}$ ``spin structures''. Each term in the sum consists of the same set of monomials but with varying signs such that after summing, all the monomials contribute with the positive sign.

In the case of the generalized W state, the graph in question is a hexagonal lattice with periodic boundary conditions.  The periodic identification implies that the genus of the graph is $1$ and so we need to sum over four terms. For our purposes, the precise form of the sum is not needed. We will simply compute the Pfaffian with a natural orientation and use the form of the partition function given by replacing the monomial coefficients with their absolute values. This would be enough for us to prove the inequality $\scg$. Our hexagonal lattice has an extra bi-partite structure because the vertices of the lattice either come from state $\psi$ (black vertex) or its conjugate $\bar\psi$ (white vertex). We will use the natural orientation coming from this bi-partition and put the arrows from $\psi$ vertex to $\bar \psi$ vertex. This gives us an orientation with which we define the anti-symmetric adjacency matrix. Among the $2m$ vertices, we will take the first $m$ to be black vertices and the next $m$ to be white vertices. The bi-partite structure means that the adjacency matrix is off-block-diagonal 
\begin{align}
    M= \begin{pmatrix}
        0 & K \\
        -K & 0 
    \end{pmatrix}
\end{align}
where $K$ is an $m\times m$ matrix of edges that connect black vertices with white vertices. For such matrix $M$, ${\rm Pf}(M)={\rm Det}(K)$. It is easy to compute the matrix $K$ and its determinant. In section \ref{mrenyi-def}, the measure ${\cal E}_n^{(\3)}$ is defined by specifying a $3$-tuple of permutation elements $(\sigma_1,\sigma_2,\sigma_3)=(g_1,g_2,{\rm id})$ that encode how the $\psi$ indices are contracted with those of $\bar \psi$'s i.e. how the black vertices are connected with white vertices. Here $g_1$ and $g_2$ are generators of ${\mathbb Z}_n\otimes {\mathbb Z}_n$. This means the matrix $K$ is the sum of tensor product matrices,
\begin{align}
    K = z_1 \,{\mathbb S}\otimes {\mathbb I}+ z_2 \,{\mathbb I}\otimes {\mathbb S} + z_3 \,{\mathbb I}\otimes {\mathbb I}
\end{align}
where ${\mathbb S}$ is the standard $n\times n$ shift matrix that is a generator of ${\mathbb Z}_n$. The matrix $K$ can be easily diagonalized by noticing that the shift matrix ${\mathbb S}$ has eigenvalues that are $n$-th roots of unity $\omega^i$, $\omega=e^{2\pi i/n}$ and $i=1,\ldots, n$. Taking the product over all the eigenvalues we get,
\begin{align}\label{detk}
    {\rm Det}(K)&=\prod_{i=1}^n\prod_{j=1}^n (z_1 \omega^i+ z_2 \omega^j + z_3)\notag\\
    &= \prod_{i=1}^{n} ((z_3+\omega^i z_2)^n- (-z_1)^n).
\end{align} 
In the last line, we used the equation $\prod_{i=1}^n(a+\omega^i b)= a^n-(-b)^n$. Even though it is not manifest, the expression \eqref{detk} is symmetric under permutations of $z_1, z_2$ and $z_3$ as expected. This is the expression that appears in equation \eqref{3partiteW} as $Z_{++}$. The correct partition function is obtained by expanding $Z_{++}$ in powers of $z_1,z_2,z_3$ and restoring the signs of all the terms to be positive. This is achieved by taking the appropriate combination of $Z_{\pm \pm}$ as in equation \eqref{3partiteW}. See \cite{KASTELEYN19611209, 10.1063/1.1703953} for details regarding this.

\bibliography{LargeDCFT}

\end{document}